\documentclass[12pt]{iopart}

\usepackage{iopams}
\usepackage{txfonts}
\usepackage{siunitx}
\usepackage{graphicx}
\usepackage{tikz}
\usepackage{url}
\usepackage{bbm}

\usepackage[english]{babel}

\usepackage{hyperref}
\usepackage{fancybox}

\begin{document}
	
\title[Cosmic Structure Formation and Fluctuation-Dissipation Relations]{Kinetic Field Theory: Cosmic Structure Formation and Fluctuation-Dissipation Relations}
	
\author{Johannes Dombrowski, Felix Fabis, Matthias Bartelmann}
\address{Universit\"at Heidelberg, Zentrum f\"ur Astronomie, Institut f\"ur Theoretische Astrophysik, Philosophenweg 12, 69120 Heidelberg, Germany}
\ead{bartelmann@uni-heidelberg.de}

\begin{abstract}
Building upon the recently developed formalism of Kinetic Field Theory (KFT) for cosmic structure formation by Bartelmann et al., we investigate a kinematic relationship between diffusion and gravitational interactions in cosmic structure formation. In the first part of this work we explain how the process of structure formation in KFT can be separated into three processes, particle diffusion, the accumulation of structure due to initial momentum correlations and interactions relative to the inertial motion of particles. We study these processes by examining the time derivative of the non-linear density power spectrum in the Born approximation. We observe that diffusion and accumulation are delicately balanced because of the Gaussian form of the initial conditions, and that the net diffusion, resulting from adding these two counteracting contributions, approaches the contributions from the interactions in amplitude over time. This hints at a kinematic relation between diffusion and interactions in KFT. Indeed, in the second part, we show that the response of the system to arbitrary gradient forces is directly related to the evolution of particle diffusion in the form of kinematic fluctuation-dissipation relations (FDRs). This result is independent of the interaction potential. We show that this relationship roots in a time-reversal symmetry of the underlying generating functional. Furthermore, our studies demonstrate how FDRs originating from purely kinematic arguments can be used in theories far from equilibrium.
\end{abstract}




\submitto{\NJP}

\section{Introduction}

Recently, Bartelmann et al. \cite{Bartelmann_main, Bartelmann_factorization, Bartelmann_Born} developed a Kinetic Field Theory (KFT) to treat cosmic structure formation based on methods introduced first by Das and Mazenko in \cite{Mazenko2010, Mazenko2011, Das2012, Das2013} and structurally similar to non-equilibrium quantum field theory. This formalism mirrors the approach of N-body simulations following particles in phase space and, thus, avoiding difficulties with shell crossing ubiquitous in conventional approaches to cosmic structure formation.

The canonical, $N$-particle ensemble considered is initially correlated in phase space and subject to the Hamiltonian equations of motion. The central object of the formalism is a generating functional containing the complete statistical information about the initial conditions and the propagation of the particles. Correlators, e.g. the density power spectrum, can be extracted from the generating functional using functional derivatives. Gravitational interactions between particles are treated perturbatively using a response function in the spirit of Martin-Siggia-Rose theory \cite{MSR_theory, MSR_Lagrangian} or can be approximated in the Born approximation \cite{Bartelmann_Born}.

In \cite{Bartelmann_main} it was demonstrated that already at first order in the interactions the non-linear power spectrum is in good agreement with $N$-body simulations down to remarkably small scales. In \cite{Bartelmann_factorization} we have shown that our formalism allows to take the full non-linear coupling of free-streaming trajectories due to initial momentum correlations into account and that the free generating functional factorizes into a single numerically tractable integral of standard form. In a separate analysis \cite{Bartelmann_Born} we show that averaging the interactions in the Born approximation allows for a computation of the non-linear power spectrum which is in remarkable agreement with $N$-body simulations with relative differences being of order $\approx \SI{15}{\percent}$ up to a wave number of $k\leq 10\,h\,\mathrm{Mpc}^{-1}$ for a scale factor of $a=1$. With the present analysis, we wish to prove that diffusion and gravitational interactions are kinematically related in KFT. 

As a first step we consider the time evolution of the density power spectrum approximating gravitational interactions in the Born approximation as in \cite{Bartelmann_Born}. One major advantage of KFT over $N$-body simulations is that we can compute analytic expressions for density correlation functions and, in this way, the study of time derivatives enables us to separate three fundamental processes in structure formation, particle diffusion, the accumulation of structure due to the initial momentum correlations and gravitational interactions. Our analysis shows that diffusion and accumulation are delicately balanced, demonstrating the eminent role of Gaussian initial conditions. We observe that the resulting net diffusion seems to be closely related to the contributions from interactions. This suggests that the processes of diffusion and interactions are kinematically related to each other. We discuss the reliability of the Born approximation for our purposes by comparing the time derivative of the non-linear power spectrum in the Born approximations with $N$-body simulations.

Motivated by this result, we show in the second part that the time evolution of particle diffusion is related to the response of the system to an arbitrary gradient force by kinematic fluctuation-dissipation relations (FDRs). This relationship is a consequence of a time-reversal symmetry of the generating functional which respects the Gaussian form of the initial conditions. Although these FDRs are purely kinematic statements, our analysis shows that they can give insight into processes far from equilibrium. To our knowledge, this is a novel application of FDRs.

This article begins with an introduction into KFT in Section \ref{sec:KFT}, which summarizes the main results from the previous works \cite{Bartelmann_main, Bartelmann_factorization, Bartelmann_Born} and introduces our notation and methods. We then study in Section \ref{sec:TD} the time derivative of the non-linear density power spectrum and discuss FDRs in Section \ref{sec:FDR}. Finally we conclude in Section \ref{sec:Conclusion} and give an outlook into future applications and relevance of FDRs for cosmic structure formation.

\section{Kinetic Field Theory for Cosmic Structure Formation}
\label{sec:KFT}

In this section we review the formalism of the kinetic field theory recently developed in \cite{Bartelmann_main} and continued in \cite{Bartelmann_factorization} and \cite{Bartelmann_Born}. This serves as an introduction to our notation and methods.

\subsection{Initially correlated Hamiltonian system}
\label{sec:Initially_correlated_Hamiltonian_system}

We study a Hamiltonian system of $N$ classical particles with identical mass, which we set to unity for simplicity. The individual particles are described by their phase-space coordinates $\vec{x}_j=(\vec{q}_j, \vec{p}_j)^{\top}$, where the index $j=1, \dots, N$ denotes the particle number. Introducing the $N$-dimensional unit vector $\vec{e}_j$ in $j$-direction, we collect the $N$ phase-space coordinates $\vec{x}_j$ into a phase-space tensor:
\begin{equation}
\mathbf{x}= \vec{x}_j\otimes\vec{e}_j,
\end{equation}
where summation over $j$ is implied. In the following, bold-faced symbols always denote tensors combining contributions from all $N$ particles. We define a scalar product between two such tenors by:
\begin{equation}
	\mathbf{a}\cdot \mathbf{b}= \vec{a}_j\cdot \vec{b}_j = \vec{a}_{q_j}\cdot\vec{b}_{q_j} + \vec{a}_{p_j}\cdot\vec{b}_{p_j}.
\end{equation}
The unit-mass particles are subject to the Hamiltonian equations of motion, which we sometimes write schematically as $\mathbf{E}(\mathbf{x})=0$. Using the linear growth factor $D_+-D_+^{(\mathrm{i})}$ as time coordinate, the Hamiltonian of our system in expanding space is given by (see \cite{Bartelmann_trajectories}):
\begin{equation}
H=\frac{\vec{p}^2(t)}{2 g}+gv,\quad g\coloneqq a^2 D_+ f H, \quad f\coloneqq \frac{\mathrm{d}\ln D_+}{\mathrm{d}\ln a}, \label{Hamiltonian_expansion}
\end{equation}
where $g$ is normalised to $1$ at the initial time, $a$ is the cosmological scale factor, $H$ is the Hubble function and $v$ is the Newtonian potential satisfying the Poisson equation
\begin{equation}
\nabla^2 v= \frac{3}{2}\frac{a}{g^2}\delta,
\end{equation}
with density contrast $\delta$.

We assume the particles to be initially correlated in phase space. Every realization $\mathbf{x}^{(\mathrm{i})}$ of the initial conditions has a probability described by a phase-space distribution $P\big(\mathbf{x}^{(\mathrm{i})}\big)$. Under the standard cosmological assumptions that the initial velocity is the gradient of a Gaussian random velocity field $\vec{p}=\vec{\nabla}\psi$ and that the initial particle distribution obeys the continuity equation $\delta=-\vec{\nabla}^2\psi$, the initial phase-space distribution is completely determined by the initial power spectrum and has the Gaussian form (see \cite{Bartelmann_main} for a careful derivation):
\begin{equation}
P\big( \mathbf{x}^{(\mathrm{i})} \big)= \frac{V^{-N}}{\sqrt{(2\pi)^{3N}\det C_{pp}}}\exp\left( -\frac{1}{2}\vec{p}_j^{\,(\mathrm{i})\top}C^{-1}_{p_j p_k} \vec{p}_k^{\,(\mathrm{i})} \right). \label{Initial_cond}
\end{equation}
The factor $\mathcal{C}(\mathbf{p}^{(\mathrm{i})})$, which additionally appears in \cite{Bartelmann_main} and describes initial density correlations, is here assumed to be unity. This is a reasonable assumption if we are interested in the late evolution of cosmic structures where the initial density correlations become subdominant compared to initial momentum correlations. The initial momentum correlations are given by the initial density power-spectrum $P_\delta(k)$:
\begin{equation}
C_{p_j p_k}\Big(\big|\vec{q}_j^{\,(\mathrm{i})}-\vec{q}_k^{\,(\mathrm{i})}\big|\Big)=\int \frac{\mathrm{d}^3k}{(2\pi)^3}\frac{\vec{k}\otimes\vec{k}}{k^4}P_{\delta}(k)\mathrm{e}^{-\mathrm{i}\vec{k}\left(\vec{q}_j^{\;\!(\mathrm{i})}-\vec{q}_k^{\;\!(\mathrm{i})}\right)}, \label{momentum_correlations}
\end{equation}
which defines the $(j,k)$-component of the matrix $C_{pp}$ appearing in \eref{Initial_cond}. The $C^{-1}_{p_jp_k}$ in \eref{Initial_cond} are then defined as the $(j,k)$-component of the inverse matrix $C^{-1}_{pp}$. Furthermore, we define $\frac{\sigma_1^2}{3}\mathbbm{1}_3\coloneqq C_{p_jp_j}$ as the initial momentum variance. 

We can already see from equations \eref{Initial_cond} and \eref{momentum_correlations} that our theory will contain two competing processes. On the one side there is particle diffusion due to the initial momentum variance $\sigma_1^2/3$. This diffusion should not be confused with thermal diffusion, but should rather be seen as an ensemble effect: The momentum of each particle seen on its own has the variance $\sigma_1^2$ when averaging over all realisations of the initial conditions. Every single realisation of the initial conditions has a completely deterministic velocity field without any local variance, i.e.\ there is no `thermal' component. Furthermore, the conditional probability $C_{p_j p_k}$ takes into account that the momenta of any two particles are not independent of each other, but depend on their initial distance $|\vec{q}_j^{\,(\mathrm{i})}-\vec{q}_k^{\,(\mathrm{i})}|$ leading to the accumulation of structure already in the free theory.

\subsection{Generating functional}

It was shown in \cite{Bartelmann_main} that the entire statistical information on the system is encoded in a generating functional in the spirit of Martin-Siggia-Rose (MSR) theory \cite{MSR_theory, MSR_Lagrangian}:
\begin{equation}
Z\big[ \mathbf{J}, \mathbf{K} \big]= \int \mathrm{d}\Gamma \int_{\mathbf{x}(0)=\mathbf{x}^{(\mathrm{i})}} \mathcal{D}\mathbf{x}\int \mathcal{D}\boldsymbol{\chi}\exp\left[ \mathrm{i}S[ \mathbf{x}, \boldsymbol{\chi} ] +\mathrm{i}\int \mathrm{d}t\left(\boldsymbol{\chi}\cdot\mathbf{K} +\mathbf{x}\cdot\mathbf{J} \right) \right], \label{gen_func}
\end{equation}
where we defined the MSR-action
\begin{equation}
S[\mathbf{x}, \boldsymbol{\chi}]\coloneqq \int\mathrm{d}t \,\boldsymbol{\chi}\cdot\mathbf{E}(\mathbf{x}). \label{MSR_action}
\end{equation}
This generating functional averages over all possible initial configurations according to the phase-space measure $\mathrm{d}\Gamma=\mathrm{d}\mathbf{x}^{(\mathrm{i})} P\big(\mathbf{x}^{(\mathrm{i})}\big)$, which represents an ensemble average. We introduced an MSR-field $\boldsymbol{\chi}$ as the Fourier conjugate to the equations of motion (integration over $\boldsymbol{\chi}$ gives a functional delta distribution ensuring that the equations of motion with an auxiliary source field $\mathbf{K}$ are satisfied). The auxiliary generator fields $\mathbf{K}$ and $\mathbf{J}$ are introduced to allow computing correlation functions through functional derivatives of the generating functional:
\begin{equation}
\left\langle \dots\mathbf{x}(t)\dots\boldsymbol{\chi}(t')\dots \right\rangle = \dots\frac{\delta}{\mathrm{i}\delta\mathbf{J}(t)}\dots\frac{\delta}{\mathrm{i}\delta\mathbf{K}(t')} \dots Z\big[ \mathbf{J}, \mathbf{K} \big]\bigg|_{\mathbf{J}=0=\mathbf{K}}. \label{Functional_derivatives}
\end{equation}
This gives a physical meaning to the field $\boldsymbol{\chi}$ as a measure for the response of the system to an external force $\mathbf{K}$. 

We define the free generating functional $Z_0\big[ \mathbf{J}, \mathbf{K} \big]$ by replacing the full equations of motion by the free equations of motion in the MSR-action \eref{MSR_action}.

If we denote the solution to the equations of motion with auxiliary source ($\mathbf{E}+\mathbf{K}=0$) by $\bar{\mathbf{x}}$, or equivalently $\bar{\mathbf{x}}_0$ in the free case, the generating functionals take the form:
\begin{equation}
Z_{0}\big[ \mathbf{J},\mathbf{K} \big] = \int\mathrm{d}\Gamma\exp\left[ \mathrm{i}\int\mathrm{d}t\,\mathbf{J}(t)\cdot \bar{\mathbf{x}}_{0}(t) \right]. \label{Generating_functional}
\end{equation}

Usually one is interested in collective observables like the particle density rather than in all the microscopic degrees of freedom described by $\mathbf{x}$. The statistical information about the density in Fourier space can be extracted from the generating functional with the density operator:
\begin{equation}
\hat{\Phi}_{\rho}\left(t_1, \vec{k}_1\right)=\sum_j \hat{\Phi}_{\rho_j}\left(t_1, \vec{k}_1\right)\coloneqq\sum_j \exp\left( -\mathrm{i}\vec{k}_1\cdot\frac{\delta}{\mathrm{i}\delta\vec{J}_{q_j}(t_1)} \right), \label{density_operator}
\end{equation}
where $\hat{\Phi}_{\rho_j}$ is a one-particle operator. In the following we abbreviate the arguments of Fourier-space operators by $1\coloneqq \left(t_1, \vec{k}_1\right)$ and $-1\coloneqq \left(t_1, -\vec{k}_1\right)$.

We also introduce a collective response field $B$ combining the microscopic response field $\boldsymbol{\chi}$ with the gradient of a density field in Fourier space ($\mathrm{i}\vec{k}_1\rho(1)$), thus describing the reaction of the system to a gradient force:
\begin{equation}
B(1)=\sum_j B_j(1)\coloneqq \mathrm{i}\vec{k}_1\cdot \vec{\chi}_{p_j}(t_1)\,\rho_j(1).
\end{equation}
The one-particle operator for this field is then given by:
\begin{equation}
\hat{\Phi}_{B_j}(1)=\left(\mathrm{i}\vec{k}_1 \cdot \frac{\delta}{\mathrm{i}\vec{K}_{p_j}(t_1)}\right)\hat{\Phi}_{\rho_j}(1) \eqqcolon \hat{b}_j(1)\hat{\Phi}_{\rho_j}(1). \label{response_operator}
\end{equation}
These collective-field operators enable us to write the full generating functional in terms of the free generating functional:
\begin{equation}
Z\big[ \mathbf{J}, \mathbf{K} \big] = \exp\left( \hat{S}_I \right) Z_0\big[ \mathbf{J}, \mathbf{K} \big], \quad \text{with} \quad \hat{S}_I=-\mathrm{i}\int\mathrm{d}1\,\hat{\Phi}_{B}(-1)v(1)\hat{\Phi}_{\rho}(1), \label{perturbation_theory}
\end{equation}
where $v$ is the Fourier-transform of the interaction potential.

\subsection{Density correlators and interactions}
\label{sec:Density_and_response_field_correlations}

The aim of KFT is to compute cosmological density correlators by applying the density operators \eref{density_operator} to the generating functional:
\begin{eqnarray}
G_{\rho\dots\rho}(1\dots n)\coloneqq \hat{\Phi}_{\rho}(1)\dots\hat{\Phi}_{\rho}(n)Z\left[ \mathbf{J}, \mathbf{K} \right]\Big|_{\mathbf{J}=0=\mathbf{K}}. \label{density_correlations}
\end{eqnarray}
The application of $n$ single-particle density operators to $Z[\mathbf{J}, \mathbf{K}]$ leads to the translation $\mathbf{J}\rightarrow \mathbf{J}+\mathbf{L}$, and thus:
\begin{equation}
\prod_{s=1}^n\hat{\Phi}_{\rho_{j_s}}Z\big[ \mathbf{J}, \mathbf{K} \big]\Big|_{\mathbf{J}=0}=Z\big[ \mathbf{L}, \mathbf{K} \big],
\end{equation}
where the shift tensor $\mathbf{L}$ is given by:
\begin{equation}
\mathbf{L}(t)= -\sum_{s=1}^n\delta_D (t-t_s)\left(\begin{array}{c}
\vec{k}_s \\
0
\end{array}\right)\otimes\vec{e}_{j_s}. \label{shift_tensor}
\end{equation}
Density correlators are thus given by $Z[\mathbf{L}, 0]$. The full generating functional is, however, not tractable in an analytic fashion. The result \eref{perturbation_theory} allows for two different ways of computing density correlators in KFT. We can either apply the density operators to the free generating functional and include the interactions by expanding the exponential in \eref{perturbation_theory}, or we approximate the full solutions $\bar{\mathbf{x}}$ appropriately and work with the full generating functional. We briefly discuss both methods in the following.

If we want to work with the free generating functional we have to define at first what we mean by `free'. If we decompose the Hamiltonian \eref{Hamiltonian_expansion} into a free and an interacting part, $H=H_0+H_I$:
\begin{equation}
H_0\coloneqq\frac{\vec{p}^2}{2}, \quad H_I\coloneqq h\frac{\vec{p}^2}{2} + gv, \quad h\coloneqq \frac{1}{g}-1, \label{splitting_Hamiltonian}
\end{equation}
then the solution to the free equation of motion with auxiliary source $\mathbf{K}$ is given by:
\begin{eqnarray}
\vec{q}_j(t)=\vec{q}_j^{(\mathrm{i})}+g_{qp}(t, 0)\vec{p}_j^{(\mathrm{i})}+ \int\mathrm{d}t'\,G^{(\mathrm{ret})}_{qp}(t, t')\vec{K}_{p_j}(t'), \label{solution_position} \\
\vec{p}_j(t)=g_{pp}(t, 0)\vec{p}_j^{(\mathrm{i})} + \int\mathrm{d}t'\,G^{(\mathrm{ret})}_{pp}(t, t')\vec{K}_{p_j}(t'), \label{solution_momentum}
\end{eqnarray}
where we neglected the $\vec{K}_{q_j}$ since they are not acted upon by the response operator \eref{response_operator}. We also defined the propagator components\footnote{In this work, the letter $G$ denotes correlation functions, see \eref{density_correlations}, as well as Green's functions, however their different index structure should prevent confusion.}:
\begin{equation}
g_{qp}(t, t')=t-t', \quad g_{pp}(t, t')=1, \quad G^{(\mathrm{ret})}_{qp/pp}(t, t')\coloneqq g_{qp/pp}(t, t')\Theta(t-t'). \label{Zel'dovich_prop}
\end{equation}
Since $t = D_+ - D_+^{(\mathrm{i})}$ this choice of `free motion' is equivalent to the Zeldovich approximation and thus contains already part of the interactions. The `remainder' of the interactions is introduced perturbatively via the interaction operator \eref{perturbation_theory}. An application of a one-particle response operator $\hat{b}_{j_r}$ on the generating functional $Z_0[\mathbf{L}, \mathbf{K}]$ gives the response factor $b_{j_r}$:
\begin{equation}
\hat{b}_{j_r}(r)Z_0\big[ \mathbf{L},\mathbf{K} \big] = \mathrm{i}\sum_{s=1}^l G^{(\mathrm{ret})}_{qp}(t_s, t_r)\vec{k}_r\vec{k}_{s}\delta_{j_r j_s} Z_0\big[ \mathbf{L},\mathbf{K} \big] \eqqcolon b_{j_r}(r)Z_0\big[ \mathbf{L},\mathbf{K} \big]. \label{response_factor}
\end{equation}
At order $m$ in the interaction operator \eref{perturbation_theory} we have to apply $m$ response operators and $m$ additional density operators. So the most general object we have to consider in this scope is the correlator of $m$ response fields and $l=m+n$ density fields computed from $Z_0[\mathbf{J}, \mathbf{K}]$:
\begin{equation}
G^{(0)}_{B_{j_1'}\dots B_{j_m'}\rho_{j_1}\dots\rho_{j_l}}(1'\dots m'1\dots l)=  \left(\prod_{r=1}^m b_{j_r}(r) \right)Z_0\big[ \mathbf{L},0 \big].
\end{equation}
It remains to compute the density correlator $Z_0[\mathbf{L}, 0]$ from \eref{Generating_functional}:
\begin{eqnarray}
Z_0[\mathbf{L}, 0]&= \int \mathrm{d}\Gamma \mathrm{e}^{\mathrm{i}\vec{L}_{q_j}\cdot\vec{q}_j^{(\mathrm{i})}+\mathrm{i}\vec{L}_{p_j}\cdot\vec{q}_j^{(\mathrm{i})}},\nonumber \\
&=V^{-N}\int\mathrm{d}\mathbf{q}^{(\mathrm{i})}\exp\left( -\frac{1}{2}\vec{L}_{p_j}^\top C_{p_j p_k}\vec{L}_{p_k} +\mathrm{i}\vec{L}_{q_j}\cdot\vec{q}_j^{\,(\mathrm{i})} \right),
\end{eqnarray}
where we introduced the spatial and momentum shift tensors:
\begin{equation}
\vec{L}_{q_j}=-\sum_s \vec{k}_s\otimes \vec{e}_{j_s}, \quad \vec{L}_{p_j}=-\sum_s g_{qp}(t_s, 0)\vec{k}_s\otimes\vec{e}_{j_s},
\end{equation}
and we integrated over the initial momenta. The momentum correlations $C_{p_jp_k}$ depend only on the relative particle separations $\vec{r}_{jk}=\vec{q}_j^{\,(\mathrm{i})}-\vec{q}_k^{\,(\mathrm{i})}$. This allows us to write $Z_0[\mathbf{L}, 0]$ in the form:
\begin{equation}
Z_0\big[ \mathbf{L},0 \big]=V^{-l}(2\pi)^3\delta_D\left(\sum_j \vec{L}_{q_j}\right)P_2(\mathbf{L}) \prod_{j}P_1 \left(L_{p_j}\right), \label{General_correlation}
\end{equation}
where we introduced the one- and two-particle factors $P_1(L_{p_j})$ and $P_2(\mathbf{L})$ as
\begin{eqnarray}
P_1\left(L_{p_j}\right)&\coloneqq\exp \left( -\frac{\sigma_1^2}{6}\vec{L}_{p_j}^2 \right), \nonumber \\
P_2(\mathbf{L})&\coloneqq\int\left(\prod_{j=2}^l\mathrm{d}^3\vec{r}_{j1}\right)\exp\left( -\frac{1}{2}\sum_{j\neq k}\vec{L}_{p_j}^\top C_{p_j p_k}\vec{L}_{p_k} +\mathrm{i}\vec{L}_{q_j}\cdot\vec{r}_{j1} \right). \label{2_particle_term}
\end{eqnarray}
As already discussed in Section \ref{sec:Initially_correlated_Hamiltonian_system} the free theory contains two competing processes, diffusion due to the initial momentum variance $\sigma_1^2/3$ of every particle seen on its own and accumulation of structure due to the conditional probability $C_{p_j p_k}$ between the momenta of different particles. This becomes more explicit in \eref{General_correlation}, where the one-particle factors $P_1\left( \vec{L}_{p_j} \right)$ describe the diffusion of particle $j$ and the two-particle factor $P_2(\mathbf{L})$ describes the accumulation of structure due to the conditional probability $C_{p_j p_k}$ between two particles.

We now discuss an alternative approach presented in detail in \cite{Bartelmann_Born} approximating the solutions to the full equations of motion in the Born approximation. The full equations of motion following from the Hamiltonian \eref{Hamiltonian_expansion} are given by:
\begin{equation}
\ddot{\vec{q}}_j=-\frac{\dot{g}}{g}\dot{q}_j-\nabla_{q_j} v. \label{equation_of_motion}
\end{equation}
We again want to write the solution of this equation in the following form:
\begin{equation}
\vec{q}_j(t)=\vec{q}_j^{(\mathrm{i})}+g_{qp}(t, 0)\vec{p}_j^{(\mathrm{i})}+ \int_0^t\mathrm{d}t' \, g_{qp}(t, t')\vec{f}_j(t'). \label{full_solution}
\end{equation}
In \cite{Bartelmann_trajectories} a propagator of the form:
\begin{equation}
g_{qp}(t, t')\coloneqq \int_{t'}^t\mathrm{d}\bar{t}\,\mathrm{e}^{h(t'')-h(t')} \label{improved_Zeldovich}
\end{equation}
was proven to be particularly useful for the study of cosmic structure formation as it contains an even larger part of the interactions than the Zel'dovich propagator. Substituting the solution \eref{full_solution} with the improved Zel'dovich propagator \eref{improved_Zeldovich} into the equation of motion \eref{equation_of_motion} shows that the force kernel $\vec{f}_j(t)$ in \eref{full_solution} has the form:
\begin{equation}
\vec{f}_j=\frac{\dot{g}}{g}h\dot{\vec{q}}_j-\nabla_{q_j} v.
\end{equation}
Substituting the solution \eref{full_solution} into the full generating functional \eref{Generating_functional} gives the density correlations:
\begin{equation}
Z[\mathbf{L}, 0]= \mathrm{e}^{-\sum_j\bar{F}_j(t) }Z_0[\mathbf{L}, 0],
\end{equation}
where
\begin{equation}
\bar{F}_j(t)\coloneqq \mathrm{i}\int_0^t\mathrm{d}t'\,\vec{L}_{p_j}(t')\cdot\vec{f}_j(t')\eqqcolon \int_0^t \mathrm{d}t'\,F_j(t, t').
\end{equation}
It was shown in \cite{Bartelmann_Born} that the function $F(t, t')\coloneqq\sum_jF_j(t, t')$ in the case of the power spectrum, i.e.\ $n=2$, can be averaged and approximated in the spirit of the Born approximation:
\begin{equation}
\langle F(t, t') \rangle=2g_{qp}(t, t')A(t')\left[ 1+k^3\int_0^\infty\frac{y^2\mathrm{d}y}{(2\pi)^2}\bar{P}_\delta(ky, t')J(y) \right],
\end{equation}
where
\begin{eqnarray}
A(t)\coloneqq h\frac{\dot{g}}{g}\frac{\dot{g}_{qp}(t, t')}{g_{qp}(t, t')}-\frac{3}{2}\frac{a}{g^2}, \nonumber \\
J(y)\coloneqq 1+\frac{1-y^2}{2y}\ln \frac{1+y}{|1-y|}, \nonumber \\
\bar{P}_{\delta}(k, t)\coloneqq \mathrm{e}^{-\frac{\sigma_1^2}{3}k^2g_{qp}^2(t, 0)}g_{qp}^2(t, 0)P_\delta^{(\mathrm{i})}(k),
\end{eqnarray}
with the initial density power spectrum $P_\delta^{(\mathrm{i})}$. In \cite{Bartelmann_Born} it is shown that an approximation of the full power spectrum in this approach is in remarkable agreement with the power spectrum from N-body simulations, with relative deviations being of order $\approx \SI{15}{\percent}$ up to a wavenumber of $k\leq 10\,h\,\mathrm{Mpc}^{-1}$.

\subsection{Factorization}
\label{sec:factorization}

We have not explained yet how $Z_0[\mathbf{L}, 0]$ can be treated. We have shown in \cite{Bartelmann_factorization} that, by introducing the internal wave vectors $\vec{k}_{ab}$ for $a>b$ and $a=3, \dots, l$, the two-particle term can be factorized into a single, numerically tractable integral of standard form:
\begin{equation}
P_2(\mathbf{L})= \left(\prod_{2\le b<a}^l\int \mathrm{d}^3\vec{k}_{ab} \right) \prod_{1\le k<j}^l\left[\Delta_{jk}\left(\vec{k}_{jk}\right)+\mathcal{P}_{jk}\left(\vec{k}_{jk}\right)\right], \label{factorization}
\end{equation}
where we defined the wave vectors $\vec{k}_{j1}$ for $j=2,\dots,l$ as:
\begin{equation}
\vec{k}_{j1}\coloneqq \vec{L}_{q_j}-\sum_{r=2}^{j-1}\vec{k}_{jr}+\sum_{r=j+1}^l\vec{k}_{rj}.
\end{equation}
Furthermore, we introduced the Dirac-delta term:
\begin{equation}
\Delta_{jk}\coloneqq (2\pi)^3 \delta_D (\vec{k}_{jk})
\end{equation}
and the function
\begin{equation}
\mathcal{P}_{jk}=\int\mathrm{d}\vec{r}_{jk}\left( \mathrm{e}^{g_{qp}^2(t, 0)k_{jk}^2\left(a_\parallel\lambda^{\parallel}+a_\perp\lambda^\perp\right)}-1 \right)\mathrm{e}^{\mathrm{i}\vec{k}_{jk}\vec{r}_{jk}}. \label{mod_power_spec}
\end{equation}
The exponent inside the parentheses is a decomposed version of the quadratic form $\vec{L}_{p_j}^\top C_{p_jp_k}\vec{L}_{p_k}$ appearing in the two-particle term \eref{2_particle_term}:
\begin{eqnarray}
\vec{L}_{p_j}^\top C_{p_jp_k}\vec{L}_{p_k}=-\vec{L}_{p_j}^\top \pi^\parallel_{jk}\vec{L}_{p_k}a_\parallel-\vec{L}_{p_j}^\top \pi^\perp_{jk}\vec{L}_{p_k}a_\perp \eqqcolon g_{qp}^2(t, 0)\vec{k}_{jk}^2\left(a_\parallel\lambda^{\parallel}+a_\perp\lambda^\perp\right),
\end{eqnarray}
where we defined $\lambda^\parallel_{jk}$ and $\lambda^\perp_{jk}$ implicitly and used the projectors parallel $\pi^\parallel_{jk}\coloneqq \hat{k}_{jk}\otimes\hat{k}_{jk}$ and perpendicular $\pi^\perp_{jk}\coloneqq \mathcal{I}_3-\pi^\parallel_{jk}$ to the unit vector $\hat{k}_{jk}$ in the direction of $\vec{k}_{jk}$.

The function $\mathcal{P}_{jk}$ acquires an intuitive meaning when considering its limit for large scales or early times ($g_{qp}^2(t, 0)k_{jk}^2\ll 1$). It was proven in \cite{Bartelmann_factorization} that, in this limit, $\mathcal{P}_{jk}$ is linear in the initial power spectrum:
\begin{equation}
\mathcal{P}_{jk}\approx -g_{qp}^2(t, 0)\lambda_{jk}^\parallel P_{\delta}^{(\mathrm{i})}\left(k_{jk}\right). \label{lin_power_spec}
\end{equation}
For the free two-point function, i.e.\ the free power spectrum, $\lambda_{21}^\parallel=-1$ showing that, in this case, $\mathcal{P}_{21}$ reduces to the linearly evolved power spectrum on large scales or early times. Thus, we can interpret $\mathcal{P}_{jk}$ as a generalization of the linearly evolved power spectrum which takes the full non-linear coupling of free trajectories by initial momentum correlations into account. Crucially, the function $\mathcal{P}_{jk}$ can be quickly evaluated numerically using e.g.\ a Levin collocation scheme \cite{Bartelmann_factorization, Levin_82, Levin_96, Levin_97}.

\section{Time derivatives of correlation functions}
\label{sec:TD}

We have seen in the last Section that in the free theory we have two physical processes governing cosmic structure formation, i.e.\ diffusion due to the initial momentum variance $\sigma_1^2/3$ and accumulation of structure due to the initial conditional probability $C_{p_jp_k}$ between the momenta of two different particles. Including gravitational interactions into the theory adds a third process to cosmic structure formation, viz.\ the mutual gravitational attraction between two particles. In this section we treat the interactions in the fashion of the Born approximation as studied in \cite{Bartelmann_Born} and summarised in the last Section. So the non-linearly evolved density power spectrum is given by:
\begin{equation}
P_{\mathrm{nl}}(t, k)= \mathrm{e}^{-\langle \bar{F}(t) \rangle}\mathrm{e}^{Q_D}\mathcal{P}_{21}, \quad \mathrm{with} \quad Q_D\coloneqq -\frac{\sigma_1^2}{3}g^2_{qp}(t, 0)k^2 \label{Born}
\end{equation}
In this Section, the propagator $g_{qp}(t, t')$ is always assumed to be the improved Zel'dovich propagator \eref{improved_Zeldovich}.

We want to understand the effect that all three processes have on cosmic structure formation. Since they appear as factors in \eref{Born}, this is best done by examining the time derivative of the non-linear power spectrum. The time derivative of the power spectrum in the Born approximation \eref{Born} is straightforward to compute:
\begin{equation}
\partial_t P_{\mathrm{nl}}(k, t) = \left( D^{(1)}_t + D^{(2)}_t + D^{(I)}_t \right) P_{\mathrm{nl}}(k, t), \label{full_td}
\end{equation}
where the operators $D^{(1)}_t$, $D^{(2)}_t$ and $D^{I}_t$ are defined as time derivative operators acting only on the 1-particle factors, i.e.\ $\mathrm{e}^{Q_D}$, the 2-particle factor, i.e.\ $\mathcal{P}_{21}$, and the interaction factor $\mathrm{e}^{-\langle \bar{F}(t) \rangle}$ respectively.

The action of these time derivative operators on the non-linear power spectrum can be computed to be:
\begin{eqnarray}
D^{(1)}_t P_{\mathrm{nl}}(k, t)&= -2 \dot{g}_{qp}(t, 0)g_{qp}(t, 0)k^2 \frac{\sigma_1^2}{3}P_{\mathrm{nl}}(k, t), \label{1_part} \\
D^{(2)}_t P_{\mathrm{nl}}(k, t)&= -2\dot{g}_{qp}(t, 0)g_{qp}(t, 0)k^2 \mathrm{e}^{Q_D-\langle \bar{F}(t) \rangle}\int_q a_\parallel\mathrm{e}^{-g_{qp}^2(t, 0)k^2a_\parallel}\mathrm{e}^{\mathrm{i}\vec{k}\vec{q}}, \label{orig_2_part} \\
D^{(I)}_t P_{\mathrm{nl}}(k, t)&= -\frac{\mathrm{d}\langle \bar{F}(t) \rangle}{\mathrm{d}t} P_{\mathrm{nl}}(k, t), \label{Interaction}
\end{eqnarray}
where the derivative of the interaction factor is given by:
\begin{equation}
\frac{\mathrm{d}\langle \bar{F}(t) \rangle}{\mathrm{d}t} = \int_0^t\mathrm{d}t'\frac{\mathrm{d}}{\mathrm{d}t}\langle F(t, t') \rangle.
\end{equation}
Note that the derivative of the boundary of the integral vanishes as $\langle F(t, t) \rangle=0$ since $g_{qp}(t, t)=0$. The derivative of the integrand $\langle F(t, t') \rangle$ does not vanish and becomes:
\begin{equation}
\frac{\mathrm{d}}{\mathrm{d}t}\langle F(t, t') \rangle= 2\dot{g}_{qp}(t, t')A(t')\left[ 1+ k^3\int_0^\infty \frac{y^2\mathrm{d}y}{(2\pi)^2}\bar{P}_\delta(ky) J(y) \right] \label{splitting_interaction}
\end{equation}
In most of the following discussion we neglect the $k$-independent part of this term as it simply describes an overall rescaling of the spectrum and we are mostly interested in $k$-dependent features. For this purpose, we introduce the notations $\hat{D}^{(I)}_t$ and $\tilde{D}^{(I)}_t$ as differential operators acting only on the $k$-dependent and $k$-independent part of $\langle F(t, t') \rangle$ respectively.

Similarly, the two-particle term \eref{orig_2_part} contains the linear evolution of the power spectrum as well as additional contributions due to the fact that we consider the full non-linear coupling of free trajectories by initial momentum correlations, see \eref{lin_power_spec} and the discussion thereafter. In the following we are mostly interested in the latter part. Thus, we define:
\begin{eqnarray}
\hat{D}^{(2)}_t P_{\mathrm{nl}}(k, t) &\coloneqq -2\dot{g}_{qp}(t, 0)g_{qp}(t, 0)k^2 \mathrm{e}^{-\langle \bar{F}(t) \rangle}\int_q a_\parallel\left(\mathrm{e}^{Q_D-g_{qp}^2(t, 0)k^2a_\parallel}-1\right)\mathrm{e}^{\mathrm{i}\vec{k}\vec{q}} \nonumber \\
&= D_t^{(2)}P_{\mathrm{nl}}(k, t)-\mathrm{e}^{-\langle \bar{F}(t) \rangle}\partial_tP_{\mathrm{lin}}(k, t), \label{2_part}
\end{eqnarray}
in this way we have subtracted the linear evolution times the Born factor $\mathrm{e}^{-\langle \bar{F}(t) \rangle}$.

\subsection{Balance between diffusion and accumulation of structure}
\label{sec:one_particle}

\begin{figure}
	\begin{indented}
		\item[]\input{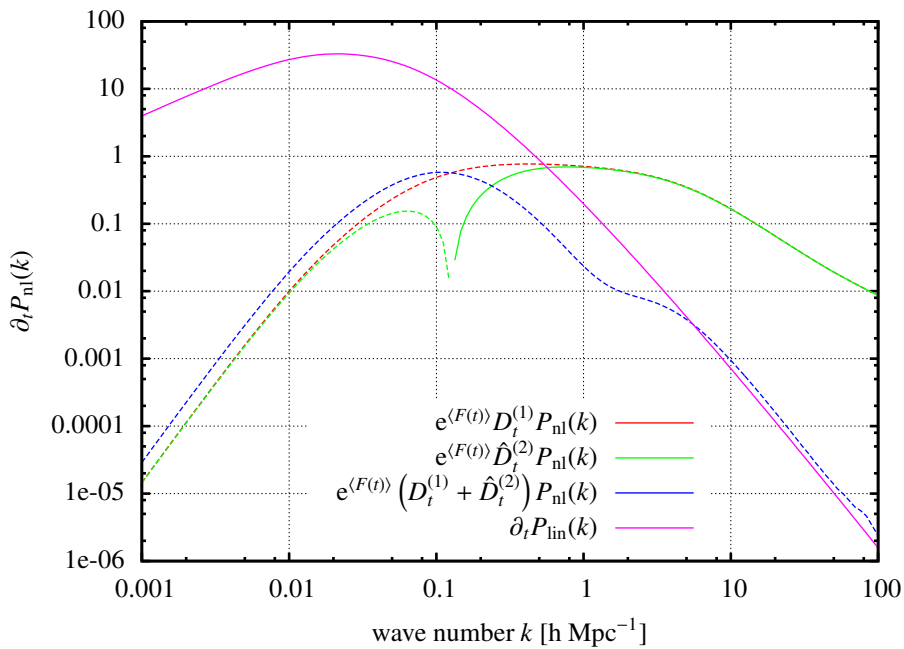}
		\caption{The 1- and 2-particle contributions to the time derivative of the power spectrum (red and green respectively) as well as their sum (blue) at a scale factor of $a=1$. Dashed lines denote negative values. The time derivative of the linearly evolved power spectrum is plotted in violet for comparison. The Born factor $\mathrm{e}^{-\langle F(t) \rangle}$ is divided out in order to make the result exact at arbitrary scales.}
		\label{fig:balance}
	\end{indented}
\end{figure}

In a first step we want to examine the one- and two-particle contributions, \eref{1_part} and \eref{2_part}, describing the diffusion and accumulation of structure due to the initial conditions. Both contributions are depicted in Fig.\ \ref{fig:balance} as well as their sum and for comparison also the time derivative of the linearly evolved power spectrum $g_{qp}^2(t, 0)P_\delta^{(\mathrm{i})}$. We divided the one- and two-particle contributions by the Born factor $\mathrm{e}^{-\langle F(t) \rangle}$ in order to make the curves independent of any shortcomings of the Born approximation on small scales. In this way the results are exact at any scale and it makes sense to depict them up to a wave number of $k= 100\, h\,\mathrm{Mpc}^{-1}$.

We observe a remarkable balance between the one- and two-particle contributions on small scales as their sum is several orders of magnitude smaller than both terms individually. This balance originates from the Gaussian form of the cosmological initial conditions and is only possible because our formalism allows to take the full non-linear coupling of free trajectories due to the initial momentum correlations into account - accounting only for part of them would lead to a strong domination of diffusion. Any small violation of this balance would lead to either very strong diffusion, thus preventing any structure formation, or much stronger structure formation than currently observed (many orders of magnitude larger than the linear evolution). An interesting consecutive question for a future study is whether this observation could constrain initial non-Gaussianities. 

\begin{figure}
	\begin{indented}
	\item[]\input{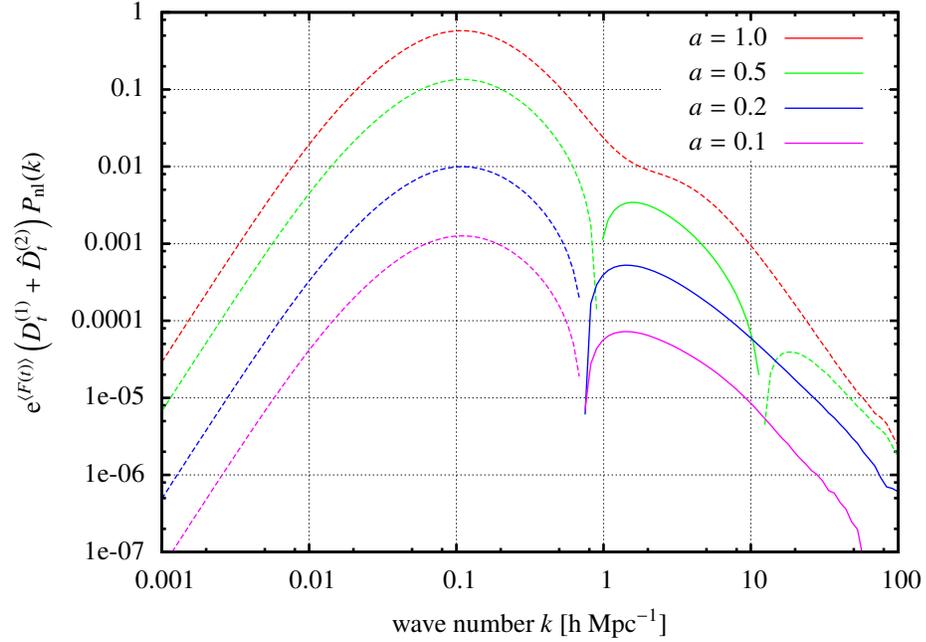}
	\caption{The sum of the 1- and 2-particle contributions at different times. Negative values are indicated by dashed lines. The Born factor $\mathrm{e}^{-\langle F(t) \rangle}$ is divided out making the results exact at arbitrary scales.}
	\label{fig:net_diffusion_evolution}
	\end{indented}
\end{figure}

The sum of the one- and two-particle contributions can be positive or negative, depending on the time we are looking at, see Fig.\ \ref{fig:net_diffusion_evolution} for its behaviour at different times. While we can see at early times that this sum leads to some structure formation on small scales, diffusion dominates at late times where it results in a net diffusion effect. 

\subsection{Interactions and net diffusion}

\begin{figure}
	\begin{indented}
	\item[]\input{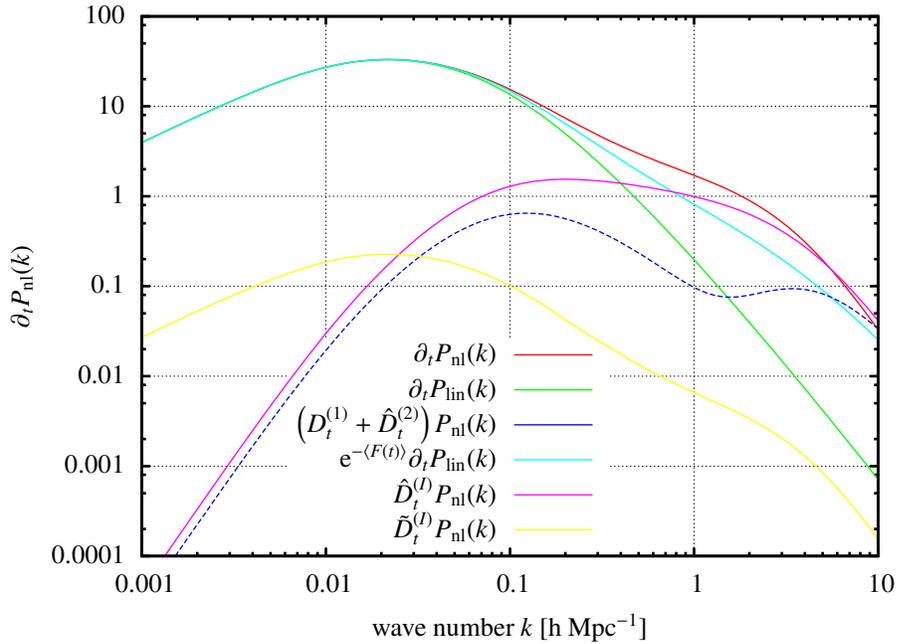}
	\caption{The time derivative of the power spectrum in the Born approximation (red) in comparison with the linear evolution (green). Furthermore, the individual contributions to the time derivative of the non-linear power spectrum are shown: net diffusion (blue), linear evolution times Born factor (light blue), derivative of the $k$-dependent (violet) and $k$-independent (yellow) part of the Born factor. Dashed curves indicate negative values.}
	\label{fig:Interaction}
	\end{indented}
\end{figure}

We now want to include the contributions from interactions \eref{Interaction} into the picture. The time derivative of the power spectrum in the Born approximation, see \eref{full_td}, is given by the net diffusion, i.e.\ the sum of the one- and two-particle contributions, now including the Born factor, the linear evolution times the Born factor, i.e.\ the part of the two-particle contributions which we neglected so far, see \eref{2_part}, and the contributions from interactions $\hat{D}^{(I)}_t P_{\mathrm{nl}}$ and $\tilde{D}^{(I)}_tP_{\mathrm{nl}}$. We depict these terms individually as well as their sum in Fig.\ \ref{fig:Interaction}. Also shown is the linear evolution for comparison. We observe that the time derivative of the $k$-independent part of the Born approximation is negligible on all scales. The time derivative of the linear power spectrum times the Born factor ensures that on large scales the linear evolution is reproduced and is also important for the non-linear evolution on small scales. The time derivative of the $k$-dependent part of the Born factor ($\hat{D}^{(I)}_tP_{\mathrm{nl}}$) follows the net diffusion closely on large scales, but has the opposite sign, on intermediate scales $\hat{D}^{(I)}_tP_{\mathrm{nl}}$ is bigger than the net diffusion and on small scales these two contributions again come quite close to each other in amplitude.

\begin{figure}
	\begin{indented}
		\item[]\input{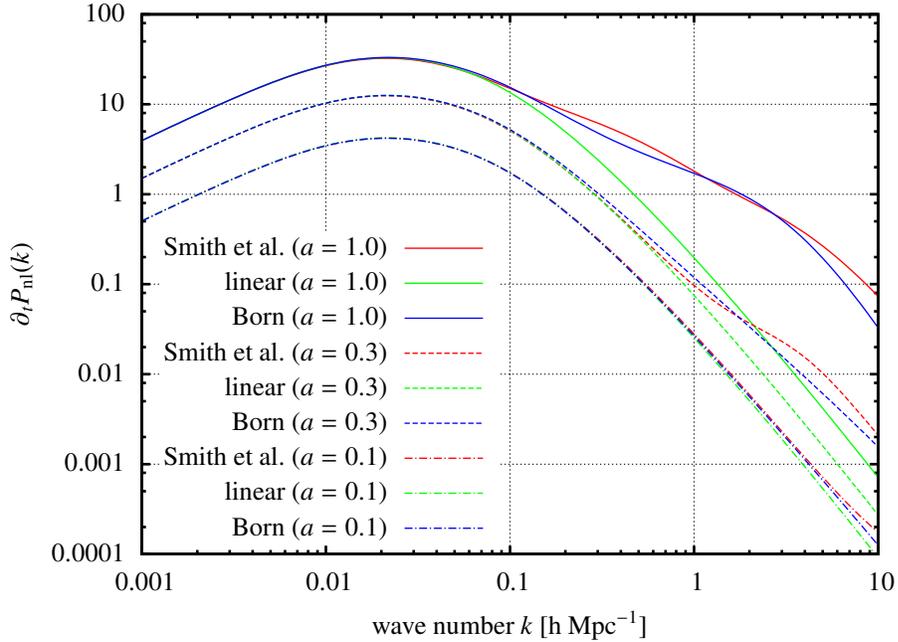}
		\caption{The time derivative of the power spectrum in the Born approximation (blue) in comparison with results from simulations \cite{Smithetal} (red). The time derivative of the linear power spectrum is plotted for comparison in green. Solid lines represent results for the scale factor $a=1$, dashed lines indicate $a=0.3$ and dashed-dotted lines $a=0.1$.}
		\label{fig:simulation}
	\end{indented}
\end{figure}

\begin{figure}
	\begin{indented}
		\item[]\input{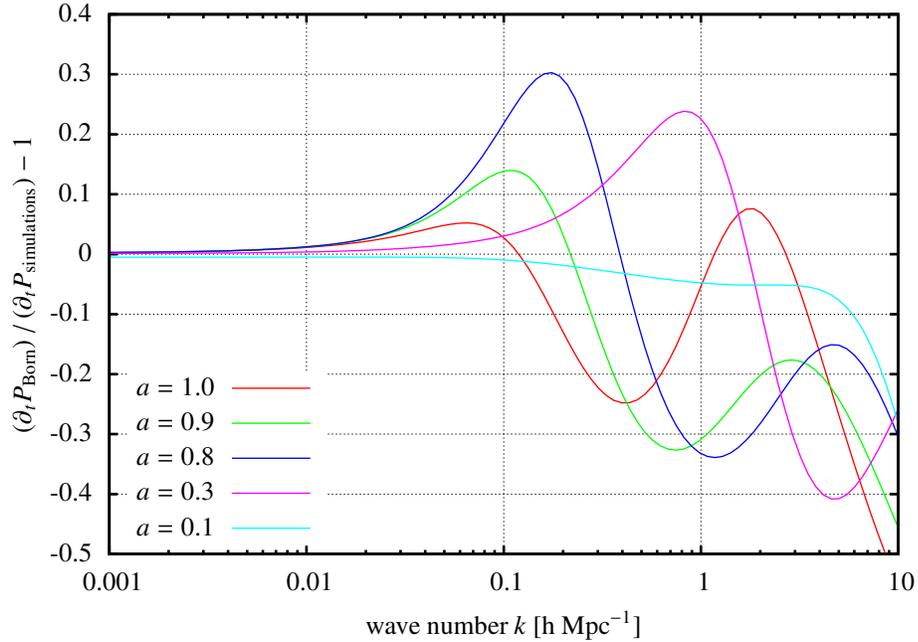}
		\caption{The relative difference between the time derivative of the power spectrum in the Born approximation and in simulations for different choices of the scale factor.}
		\label{fig:simulation_relative}
	\end{indented}
\end{figure}

Before we can draw any further conclusions, it is important to check the reliability of the Born approximation by comparing our results with the time derivative of the power spectrum obtained from $N$-body simulations \cite{Smithetal}. We depict the result from the Born approximation \eref{full_td} in comparison to $N$-body simulations in Fig.\ \ref{fig:simulation} for three choices of the scale factor $a=1,\,0.3, \,0.1$. In Fig.\ \ref{fig:simulation_relative} we plot the relative difference between our results and $N$-body simulations for the same choices of the scale factor as well as the scales factors $a=0.9,\,0.8$ which we discuss in more detail later on.

At first we discuss the results for the scale factor $a=1$ in Figs.\ \ref{fig:simulation} and \ref{fig:simulation_relative}. When comparing the results in the Born approximation with simulations, we observe a qualitatively similar behaviour as in \cite{Bartelmann_Born} where the non-linear power spectrum \eref{Born} was compared with simulations. The Born approximation is able to describe the non-linear structures up to a scale of $k\sim 5\, h\,\mathrm{Mpc}^{-1}$ to remarkable accuracy with the absolute value of the relative error being on average of the order of $\sim \SI{15}{\percent}$ and never bigger than $\SI{25}{\percent}$. On scales beyond $k\sim 5\, h\,\mathrm{Mpc}^{-1}$ the Born approximation falls strongly below the results from simulations.

Considering the evolution of the relative error in time we see that, as expected, the Born approximation predicts the early evolution ($a=0.1$) reasonably well. However, the relative error is of order $\sim \SI{30}{\percent}$ for scale factors between $a=0.3$ and $a=0.9$ and on large scales the error does not increase monotonically with redshift. The curves for $a=0.3$ in Fig.\ \ref{fig:simulation} show that the Born approximation fails to predict the correct form for the onset of the non-linear structure at $1\, h\,\mathrm{Mpc}^{-1}\lesssim k\lesssim 5\, h\,\mathrm{Mpc}^{-1}$. Notice that for $k\gtrsim 5\, h\,\mathrm{Mpc}^{-1}$ and $a=0.3$ the magnitude of the relative error decreases again. This contradicts the intuition that a perturbative theory of structure formation should be accurate on linear and mildly non-linear scales, but should fail on highly non-linear scales. However, the Born approximation is a non-perturbative, but approximate approach to cosmic structure formation and it is not clear whether this intuition is reasonable in this case. The Born approximation seems to be especially well suited for the study of the late time non-linear evolution. The time evolution of the relative error needs to be studied in detail together with the reliability and limitations of the Born approximation in a future analysis.

For our purposes here it is merely important that the Born approximation at late times ($a=1, \,0.9, \,0.8$) is in agreement with simulations at a level of $\lesssim\SI{30}{\percent}$ on scales up to $k\sim 5\, h\,\mathrm{Mpc}^{-1}$.

\begin{figure}
	\begin{indented}
	\item[]\input{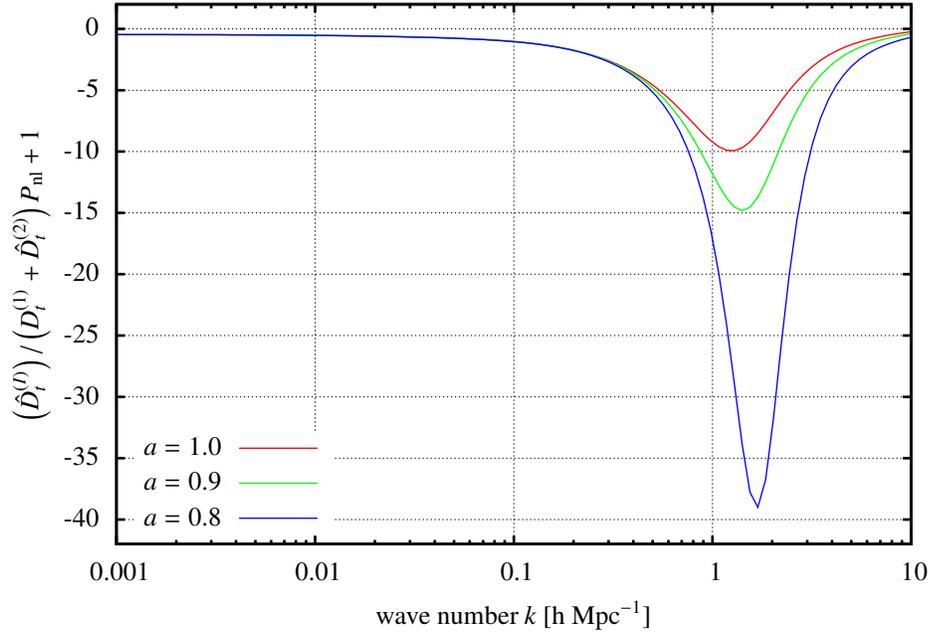}
	\caption{The sum of the interaction term $\hat{D}^{(I)}_tP_{\mathrm{nl}}$ and the net diffusion $\left( D^{(1)}_t+\hat{D}^{(2)}_t \right)P_{\mathrm{nl}}$ relative to the net diffusion for different scale factors.}
	\label{fig:equilibrium}
	\end{indented}
\end{figure}

With these caveats in mind we want to analyse the processes of net diffusion $\left( D^{(1)}_t+\hat{D}^{(2)}_t \right)P_{\mathrm{nl}}$ and the $k$-dependent part of the interactions $\hat{D}^{(I)}_t$ more closely. We plot the sum of both contributions relative to the net diffusion in Fig.\ \ref{fig:equilibrium}. We observe that the two contributions become significantly closer in amplitude for late times. Notice that this tendency of the two terms to approach each other in amplitude is most notably visible on scales $k< 5\, h\,\mathrm{Mpc}^{-1}$ where the Born approximation is in reasonably good agreement with simulations. On large scales the two contributions seem to follow a close relationship.

These observations would need to be considered unnatural unless there is some mechanism relating the amplitude of both processes to each other. This motivates our analysis in the next Section where we indeed find a relation between interactions and diffusion in terms of fluctuation-dissipation relations.

\section{Fluctuation-Dissipation Relations}
\label{sec:FDR}

In this Section we prove a fundamental connection between diffusion and interactions in terms of fluctuation-dissipation relations. We show that FDRs in KFT relate the process of diffusion with the reaction of the free system to an arbitrary gradient force. In this Section we work in the `free' theory, where the trajectories of free particles are given by \eref{solution_position} and \eref{solution_momentum} for $\mathbf{K}=0$, the propagator is given by the Zel'dovich propagator \eref{Zel'dovich_prop} and the reaction to the system to an arbitrary force $\mathbf{K}$ is computed via the response operator \eref{response_operator}. This makes the relations independent of the interaction potential.

\subsection{Density autocorrelation}
\label{FDT_autocorrelation}

In a first step we examine the single-particle density autocorrelation $G^{(0)}_{\rho_{j}\rho_j}(12)$ in more detail. In this case the system is effectively a one-particle system and the initial probability distribution \eref{Initial_cond} reduces to a Maxwellian velocity distribution:
\begin{equation}
P\left( \vec{q}_j^{\,(\mathrm{i})}, \vec{p}_j^{\,(\mathrm{i})} \right)= \frac{V^{-1}}{\left(2\pi \frac{\sigma_1^2}{3}\right)^{3/2}}\exp \left( -\frac{3}{2\sigma_1^2}\vec{p}_j^{\,(\mathrm{i})2} \right).
\end{equation}
Thus, the autocorrelation is purely described by the dissipative one-particle part $P_1(L_{p_j})$ and two-particle factors are absent:
\begin{equation}
G^{(0)}_{\rho_{j}\rho_j}(12)=V^{-1}(2\pi)^3\delta_D\left(\vec{k}_1+\vec{k_2}\right)P_1\left(L_{p_j}\right).
\end{equation}
The Dirac $\delta$-distribution reflects the homogeneity of the system and leads to the following form of the momentum shift vector $\vec{L}_{p_1}$:
\begin{equation}
\vec{L}_{p_j}= -\vec{k}_1g_{qp}(t_1, 0)-\vec{k}_2g_{qp}(t_2, 0)=-\vec{k}_1 g_{qp}(t_1, t_2)=-\vec{k}_2g_{qp}(t_2, t_1), \label{auto_momentum_shift}
\end{equation}
which is invariant under time-translation in the case of Zel'dovich trajectories. Having determined $\vec{L}_{p_j}$, the time derivative of the autocorrelation can be computed:
\begin{equation}
\frac{\partial}{\partial t_1}G^{(0)}_{\rho_{j}\rho_{j}}(12) = \frac{\sigma_1^2}{3}\vec{k}_1^2 g_{qp}(t_2, t_1) G^{(0)}_{\rho_{j}  \rho_{j}}(12), \label{td_auto_corr}
\end{equation}
where we used $\partial_{t_1}g_{qp}(t_2, t_1)=-1$ for the Zel'dovich propagator. 

This expression for the time derivative of the autocorrelation becomes most interesting when comparing it with the response function:
\begin{equation}
G^{(0)}_{B_j\rho_j}(12)= b_{j_1}(1)G^{(0)}_{\rho_{j}\rho_{j}}(12)= -\mathrm{i}G^{(\mathrm{ret})}_{qp}(t_2, t_1)\vec{k}_1^2 G^{(0)}_{\rho_{j}\rho_{j}}(12). \label{auto_corr_response}
\end{equation}
We can conclude:
\begin{equation}
-\mathrm{i}G^{(0)}_{B_j\rho_j}(12)=-\frac{3}{\sigma_1^2} \Theta(t_2 -t_1) \frac{\partial}{\partial t_1}G^{(0)}_{\rho_{j}\rho_{j}}(12). \label{fdt}
\end{equation}
This relation between the response function and the time derivative of the autocorrelation has exactly the form of a fluctuation-dissipation relation.

We have multiplied the response field on the left-hand side of \eref{fdt} with $-\mathrm{i}$ because this is the correctly normalized response field measuring the reaction of the system to two-particle interactions as can be seen in \eref{perturbation_theory}, where this factor also appears.

Fluctuation-dissipation relations are known to hold for many statistical systems \cite{FDT_review}. For example, in classical linear response theory one considers a system in thermal equilibrium and examines the response of an observable $\mathcal{O}$ to a small perturbation away from equilibrium. This is measured by the response function $\chi(t, t')$. Using time-translation invariance in thermal equilibrium, the standard textbook result can be derived:
\begin{equation}
\chi(t, t')=-\beta\Theta(t-t')\frac{\partial}{\partial t}\left\langle \mathcal{O}(t)\mathcal{O}(t') \right\rangle_{\mathrm{eq}}. \label{Kubo}
\end{equation}
This result is remarkable as it tells us that the response of the system to a small perturbation is the same as the evolution of equilibrium correlations. Examination of the equilibrium systems thus gives information on non-equilibrium physics. $\beta$ is here the inverse temperature, so comparison of \eref{fdt} and \eref{Kubo} shows that in KFT $\sigma_1^2/3$ assumes the role of the temperature.

The FDR in KFT \eref{fdt} has the same mathematical form as the conventional FDR \eref{Kubo}. However, there are some significant physical differences between the relations since KFT is a theory far from equilibrium. We discuss this in Section \ref{sec:kinematic_fdr} in more detail, but at first we want to generalise the result \eref{fdt}.

\subsection{General density correlators}
\label{sec:FDT_general}

Time translation invariance (TTI) is an important requirement for the validity of FDRs. In the case of the autocorrelation, TTI follows from homogeneity allowing us to write $\vec{L}_{p_j}$ in the form \eref{auto_momentum_shift}. If the structure of the correlator is not that simple, time-translation invariance might not hold for all momentum shift vectors $\vec{L}_{p_j}$. This leads to deviations from the FDR in its conventional form.

The most general density correlator $Z_0[\mathbf{L}, 0]$ has an arbitrary number $n_j$ of density fields with particle index $j$ and equivalently for any other particle index:
\begin{equation}
Z_0[\mathbf{L}, 0]=G^{(0)}_{\rho_j\dots\rho_j\rho_k\dots\rho_k\dots}(1\dots n_j1'\dots n_k'\dots)\eqqcolon G^{(0)}_{\rho_j\dots\rho_j\dots}(1\dots n_j\dots).
\end{equation}
We introduced the last term as an abbreviation which we use in the following. For this correlator, the momentum shift vector $\vec{L}_{p_j}$ has the form:
\begin{equation}
\vec{L}_{p_j}=-\sum_{s=1}^{n_j}\vec{k}_sg_{qp}(t_s, 0),
\end{equation}
where $n_j$ is the number of applied density operators with particle index $j$. We split $\vec{L}_{p_j}$ into a part respecting TTI and a part which breaks this invariance:
\begin{equation}
\vec{L}_{p_j} = -\sum_{s=2}^{n_j}\vec{k}_sg_{qp}(t_s, t_1) +\Delta\vec{L}_{p_j}(t_1),
\end{equation}
where the time variable $t_1$ is arbitrary so far and the TTI breaking part of $\vec{L}_{p_j}$ is given by:
\begin{equation}
\Delta\vec{L}_{p_j}(t_1)= -\sum_{s=1}^{n_j}\vec{k}_sg_{qp}(t_1, 0). \label{TTIbreaking}
\end{equation}
We see that $\Delta\vec{L}_{p_j}$ vanishes for any autocorrelation, where $n_j$ is identical with the full number of applied density operators and statistical homogeneity thus ensures $\sum_{s=1}^{n_j}\vec{k}_s=0$. In the general case, however, $\Delta\vec{L}_{p_j}$ is non-zero and leads to a correction term in the fluctuation-dissipation relation. Analogous to the steps \eref{td_auto_corr} to \eref{fdt} we derive the relation:
\begin{equation}
-\mathrm{i}b_{j}(1)P_1\left(L_{p_j}\right)=-\left(\frac{3}{\sigma_1^2}\partial_{t_1} -\mathrm{i}\Delta\vec{L}_{p_j}(t_1)\vec{k}_1 \right)P_1\left(L_{p_j}\right),
\end{equation}
where we assumed $t_s>t_1$ for all $1<s\leq n_j$ and $t_1$ is now the time where we evaluate the response function. We see that breaking TTI leads to a violation of the FDR in its conventional form. The response factor on the left hand side measures the propagation of an inhomogeneity from $t_1$ to the times $t_s$. However, $P_1\left(L_{p_j}\right)$ describes the diffusion from the initial time to the times $t_s$. Only if the modes satisfy $\sum_{s=1}^{n_j}\vec{k}_s=0$, can we neglect the propagation of particle $j$ from the initial time to $t_1$. Thus, in general, we have to subtract the diffusion taking place between the initial time and $t_1$, which is encoded in $\Delta\vec{L}_{p_j}(t_1)$.

In order to quantify the diffusion taking place between $t_1$ and the times $t_s$, we introduce a covariant time derivative being invariant under time translation:
\begin{equation}
\mathcal{D}_{t_1}^{(1, j)}\coloneqq D_{t_1}^{(1, j)}-\frac{\sigma_1^2}{3}\Delta\vec{L}_{p_j}(t_1)\vec{k}_1,
\end{equation}
where the operator $D_t^{(1, j)}$ is defined as a time derivative acting only on the one-particle factor of particle $j$, i.e.\ $P_1\left( \vec{L}_{p_j} \right)$. In terms of this time derivative, the FDR for a general density correlator, assuming $t_s>t_1$ for all $1<s\leq n_j$, has the form:
\begin{equation}
-\mathrm{i}G^{(0)}_{B_j\rho_j\dots\rho_j\dots}(12\dots n_j\dots)=-\frac{3}{\sigma_1^2}\mathcal{D}_{t_1}^{(1, j)}G^{(0)}_{\rho_j\dots\rho_j\dots}(1\dots n_j\dots). \label{mod_FDR}
\end{equation}

We see that the diffusion of particle $j$ and the reaction of particle $j$ to an arbitrary external force are kinematically related and this result is not restricted to the autocorrelation \eref{fdt}, but is valid for an arbitrary correlator if we substitute the full time derivative by the covariant time derivative $\mathcal{D}_t^{(1, j)}$ describing the evolution of diffusion.

In the following we show that this connection between the one-particle ensemble diffusion and the response function goes even deeper and is a consequence of the structure of the generating functional and the Gaussian form of the cosmological initial conditions.

\subsection{Time-Reversal Symmetry}

In statistical field theories, FDRs are typically connected to a time-reversal symmetry of the generating functional \cite{Andreanov}. We will show that the same is the case for our kinetic field theory. This gives an alternative way of deriving the FDRs considered above and leads to an easy generalization to higher-order response functions, which will turn out to be related to higher-order time derivatives.

First of all we note that the formalism of KFT is originally designed for times later than the initial time $t=0$. Hence it is not immediately clear what a time-reversal symmetry means within this formalism. However, in principle there is no need to restrict the formalism to positive times. The solutions to the equations of motion \eref{solution_position} and \eref{solution_momentum} are also valid for negative times. Thus, we can propagate the particles from their initial conditions backwards in time and calculate correlation functions at negative times by functional derivatives of the generating functional $Z[\mathbf{J}, \mathbf{K}]$ with respect to the auxiliary fields at negative times:
\begin{equation}
\left\langle \dots\mathbf{x}(-t)\dots\boldsymbol{\chi}(-t')\dots \right\rangle = \dots\frac{\delta}{\mathrm{i}\delta\mathbf{J}(-t)}\dots\frac{\delta}{\mathrm{i}\delta\mathbf{K}(-t')} \dots Z\big[ \mathbf{J}, \mathbf{K} \big]\bigg|_{\mathbf{J}=0=\mathbf{K}},
\end{equation}
where $t, t'>0$.

Our aim is to find a transformation $\mathcal{T}$ of the microscopic fields $\mathbf{x}$ and $\boldsymbol{\chi}$ which reverses the time coordinate and leaves the generating functional invariant. The time-reversed generating functional is defined as:
\begin{equation}
\mathcal{T}Z\big[ \mathbf{J}, \mathbf{K} \big]\coloneqq \int \mathrm{d}\Gamma \int \mathcal{D}\mathbf{x}\int \mathcal{D}\boldsymbol{\chi}\exp\bigg[ \mathrm{i}\mathcal{T}S[\mathbf{x}, \boldsymbol{\chi}] + \mathrm{i}\int \mathrm{d}t\,\Big( \boldsymbol{\chi}\cdot\mathbf{K} +\mathbf{x}\cdot\mathbf{J} \Big) \bigg] \label{TR_generating_func}
\end{equation}
and in this way contains the statistical information of the system with time-reversed dynamics. The phase-factor with the auxiliary fields is not transformed. Thus, correlation functions are computed in the same way as before, but they are now evaluated with time-reversed dynamics.

If the generating functional is invariant under the transformation $\mathcal{T}$, any correlation function will be invariant as well. This can be shown by substituting $\mathbf{x}\rightarrow\mathcal{T}\mathbf{x}$ and $\boldsymbol{\chi}\rightarrow\mathcal{T}\boldsymbol{\chi}$ in the path integrals in \eref{TR_generating_func} and using that a time-reversal symmetry has to be its own inverse:
\begin{eqnarray}
\fl\mathcal{T}Z\big[ \mathbf{J}, \mathbf{K} \big]&=& \int \mathrm{d}\Gamma \int \mathcal{D}(\mathcal{T}\mathbf{x})\int \mathcal{D}(\mathcal{T}\boldsymbol{\chi}) \exp\left[ \mathrm{i}\mathcal{T}S[\mathbf{x}, \boldsymbol{\chi}] + \mathrm{i}\int \mathrm{d}t\,\mathcal{T}\mathcal{T}\left( \boldsymbol{\chi}\cdot\mathbf{K} +\mathbf{x}\cdot\mathbf{J} \right) \right] \nonumber \\
\fl&=&\int \mathrm{d}\Gamma \int \mathcal{D}\mathbf{x}\int \mathcal{D}\boldsymbol{\chi}\exp\left[ \mathrm{i}S[\mathbf{x}, \boldsymbol{\chi}] + \mathrm{i}\int \mathrm{d}t\,\left( \left(\mathcal{T}\boldsymbol{\chi}\right)\cdot\mathbf{K} +\left(\mathcal{T}\mathbf{x}\right)\cdot\mathbf{J} \right) \right].
\end{eqnarray}
In this form, we see that functional derivatives of the time-reversed generating functional give time-reversed correlation functions, but with the original dynamics. Thus, invariance of the generating functional also shows invariance of correlation functions under time reversal.

We now aim at finding the explicit form of $\mathcal{T}$. Since the density and response field operators, \eref{density_operator} and \eref{response_operator}, act only on the spatial part of $\mathbf{J}$ and the momentum part of $\mathbf{K}$, we can restrict ourselves here to the case where the momentum part of $\mathbf{J}$ and the spatial part of $\mathbf{K}$ vanish. Using time-reversal invariance of the Hamiltonian equations of motion, we prove in \ref{Time_reversal_prove} that the following transformation is a time-reversal symmetry of the free generating functional:
\begin{eqnarray}
\mathcal{T}: \left\{\begin{array}{ll}
\vec{q}_j(t) &\rightarrow \;\; \vec{q}_j(-t), \\
\vec{p}_j(t) &\rightarrow \;\; -\vec{p}_j(-t), \\
\vec{\chi}_{q_j}(t) &\rightarrow \;\; -\vec{\chi}_{q_j}(-t), \\
\vec{\chi}_{p_j}(t) &\rightarrow \;\; \vec{\chi}_{p_j}(-t) -\mathrm{i}C^{-1}_{p_j p_k} \vec{p}_k(-t) -\vec{c}_j[\mathbf{J}, -t],
\end{array}\right. \label{TR_symmetry}
\end{eqnarray}
where the term $\vec{c}_j[\mathbf{J}, t]$ is a functional of $\mathbf{J}$ and is defined as: 
\begin{equation}
\vec{c}_j[\mathbf{J}, t]=g_{qp}(t, 0)\int_{-\infty}^{\infty}\mathrm{d}t'\,\vec{J}_{q_j}(t'). \label{c_term}
\end{equation}
It describes the breaking of time-translation invariance similar to the term $\Delta\vec{L}_{p_j}$ in \eref{TTIbreaking}. Indeed, if we substitute $\mathbf{J}\rightarrow\mathbf{L}$ in \eref{c_term}, as is the case for density correlators, \eref{TTIbreaking} and \eref{c_term} become completely equivalent.

The transformation laws for the microscopic degrees of freedom allow us to derive transformation laws for the macroscopic fields $B$ and $\rho$. Defining $\tilde{1}\coloneqq \left(-t_1, \vec{k}_1\right)$, the time-reversed density and response fields become:
\begin{eqnarray}
\mathcal{T}\rho_j(1)&=\rho_j(\tilde{1}), \label{density_symm} \\
\mathcal{T}B_j(1)&=B_j(\tilde{1})+\Delta B_j(\tilde{1}), \label{collective_symm}
\end{eqnarray}
with
\begin{equation}
\Delta B_j(\tilde{1})\coloneqq\vec{k}_1C_{p_jp_k}^{-1}\vec{p}_k(-t_1)\rho_j(\tilde{1})-\vec{k}_1\cdot\vec{c}_j[\mathbf{J}, -t_1]\rho_j(\tilde{1}).
\end{equation}
The first term on the right-hand side has a simple interpretation if we consider this term to be embedded into a density correlation. We prove in \ref{sec:one_particle_time_derivative} that:
\begin{equation}
\left\langle \vec{k}_1C_{p_jp_k}^{-1}\vec{p}_k(t_1)\rho_j(1)\dots \right\rangle = \mathrm{i}\frac{3}{\sigma_1^2}D_{t_1}^{(1,j)} Z_0 \big[ \mathbf{L}, 0 \big], \label{one_particle_derivative}
\end{equation}
where the dots denote an arbitrary number of further density fields. Thus, we can interpret the term $\Delta B_j(\tilde{1})$ as the evolution of time-reversed diffusion:
\begin{equation}
\left\langle \Delta B_j(\tilde{1}) \dots \right\rangle = \mathrm{i}\frac{3}{\sigma_1^2}\mathcal{D}_{-t_1}^{(1, j)} Z_0 \big[ \mathbf{L}, 0 \big]. \label{time_reversed_diffusion}
\end{equation}
In the following two paragraphs we want to build up some intuition for this term. 

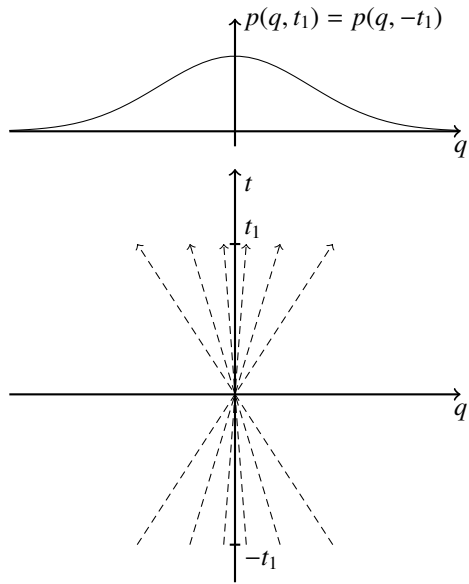
\begin{figure}
\begin{indented}
\item[]\begin{center}
\begin{tikzpicture}
\draw [->, densely dashed] (-0.6,-2) -- (0.6,2);
\draw [->, densely dashed] (0.6,-2) -- (-0.6,2);
\draw [->, densely dashed] (-0.15,-2) -- (0.15,2);
\draw [->, densely dashed] (0.15,-2) -- (-0.15,2);
\draw [->, densely dashed] (-1.3,-2) -- (1.3,2);
\draw [->, densely dashed] (1.3,-2) -- (-1.3,2);
\draw [->, thick] (-3,0) -- (3,0);
\draw [->, thick] (0,-2.5) -- (0,3);
\draw [domain=-3:3, samples=50] plot (\x, {3.5+exp(-\x*\x/2)});
\node[right](time) at (0,2.8) {$t$};
\node[below](position) at (3,0) {$q$};
\draw[thick] (-0.07,2) -- (0.07,2);
\node[right](t1) at (0,2.2) {$t_1$};
\draw[->, thick] (-3,3.5) -- (3,3.5);
\draw[->, thick] (0,3.3) -- (0,5);
\node[below](q2) at (3,3.5) {$q$};
\node[right](prob) at (0,5) {$p(q, t_1)=p(q, -t_1)$};
\draw[thick] (-0.07,-2) -- (0.07,-2);
\node[right](-t1) at (0,-2.2) {$-t_1$};
\end{tikzpicture}
\end{center}
\end{indented}
\caption{Sketch of different realizations (dashed lines) of the trajectory of one particle with fixed initial position propagated also to negative times backwards from the initial conditions. The probability $p(q, t_1)$ to find the particle at position $q$ at time $t_1$ is Gaussian and symmetric under time reversal.}
\label{fig:trajectories}
\end{figure}

A time-reversal symmetry relates the probability of a path and its time-reversed version. To illustrate this, we consider a single particle of the ensemble and choose one realization of its initial position $\vec{q}$, but leave the initial momentum free. The trajectories of the different realizations of the momentum form a double cone as shown in Figure \ref{fig:trajectories} if we propagate the particle also backward in time. At each time the probability to find the particle at some position $\vec{q}$ is given by a Gaussian with variance $\sigma_1^2/3g_{qp}^2(t, 0)$, which is symmetric in time reversal. This symmetry is the reason why density correlations are invariant under time reversal, see \eref{density_symm}.

However, we see in Figure \ref{fig:trajectories} that for positive times diffusion takes place, while for negative times the process of diffusion is reversed. The response field, being the functional Fourier conjugate of the equations of motion, see \eref{MSR_action}, is sensitive to the direction of time. Thus, the response field is not invariant under time reversal, but the time-reversed diffusion has to be taken into account in the transformation \eref{collective_symm}.

The form \eref{collective_symm} of the symmetry together with \eref{time_reversed_diffusion} is reassuringly similar to the time-reversal symmetries known from statistical field theory. For example Andreanov et al. \cite{Andreanov} consider the Langevin dynamics:
\begin{equation}
\partial_t X(t)=-\nabla V\big( X(t) \big) +\eta(t), \label{Langevin_dyn}
\end{equation}
with potential $V(X)$ and where the stochastic force $\eta(t)$ has variance $2T$. They find the time-reversal symmetry:
\begin{eqnarray}
\mathcal{T}X(t)&=X(-t), \\
\mathcal{T}\hat{X}(t)&= \hat{X}(-t)-\frac{1}{T}\partial_{-t}X(-t), \label{Langevin_symm}
\end{eqnarray}
where $\hat{X}$ takes on the role of the MSR-response function equivalent to $\boldsymbol{\chi}$. Comparing with \eref{collective_symm} and \eref{time_reversed_diffusion}, we see that the temperature plays the same role as the velocity variance $\sigma_1^2/3$. The second term on the right-hand side in \eref{Langevin_symm} describes diffusion analogous to \eref{time_reversed_diffusion}.

\subsection{Fluctuation-dissipation relations from the time-reversal symmetry}

In statistical field theory, the time-reversal symmetry can be used to derive FDRs. For the Langevin dynamics \eref{Langevin_dyn} invariance of correlations under the symmetry \eref{Langevin_symm} gives \cite{Andreanov}:
\begin{equation}
\left\langle \hat{X}(t')X(t) \right\rangle = \mathcal{T}\left\langle \hat{X}(t')X(t) \right\rangle= \left\langle \hat{X}(t)X(t') \right\rangle-T^{-1}\partial_{t}\Big\langle X(t)X(t') \Big\rangle.
\end{equation}
Due to causality the first term on the right-hand side has to vanish if $t>t'$ and we arrive at the fluctuation-dissipation relation:
\begin{equation}
\left\langle \hat{X}(t')X(t) \right\rangle =-T^{-1}\Theta(t-t')\partial_{t}\Big\langle X(t)X(t') \Big\rangle.
\end{equation}

In the following we aim at a similar derivation of the FDRs for KFT using the time-reversal symmetry \eref{collective_symm}. Since the transformation $\mathcal{T}$ leaves correlation functions invariant, we can conclude for example:
\begin{equation}
G^{(0)}_{B_j\rho_j}(\tilde{1}\tilde{2})=\mathcal{T}\left(G^{(0)}_{B_j\rho_j}(\tilde{1}\tilde{2}) \right) = G^{(0)}_{B_j\rho_j}(12) + \mathrm{i}\frac{3}{\sigma_1^2} \mathcal{D}^{(1,j)}_{t_1}G^{(0)}_{\rho_j\rho_j}(12).
\end{equation}
For $t_2>t_1$ the left hand-side has to vanish due to causality. Using that the operator $\mathcal{D}^{(1, j)}_{t_1}$ is equivalent to a regular time derivative for the auto-correlation, we arrive at the FDR \eref{fdt}:
\begin{equation}
G^{(0)}_{B_j\rho_j}(12) =-\mathrm{i}\frac{3}{\sigma_1^2}\Theta(t_2-t_1) \partial_{t_1}G^{(0)}_{\rho_j\rho_j}(12).
\end{equation}

The time-reversal symmetry not only reproduces the FDR \eref{fdt}, but also gives rise to a whole hierarchy of relations between response functions of arbitrary order and time derivatives of correlation functions. Most generally, the time-reversal symmetry gives us:
\begin{eqnarray}
&\left\langle B_{j_1'}(1')\dots B_{j_m'}(m')\rho_{j_{1}}(1) \dots \rho_{j_{n}}(n) \right\rangle_0 \nonumber \\
&= \left\langle \left(B_{j_1'}(\tilde{1'})+ \Delta B_{j_1'}(\tilde{1'}) \right)\dots \left(B_{j_m'}(\tilde{m'})+ \Delta B_{j_m'}(\tilde{m'}) \right)\rho_{j_{1}}(\tilde{1}) \dots \rho_{j_{n}}(\tilde{n}) \right\rangle_0. \label{TR:master_equation}
\end{eqnarray}
This proves that response functions of arbitrary order are related to diffusion.

\subsection{Kinematic FDRs far from equilibrium}
\label{sec:kinematic_fdr}

Since KFT is a theory far from equilibrium, there are some significant physical differences between FDRs in KFT and in thermal systems.

The physical picture behind the FDR in a thermal system is the following. Particles in a thermal system necessarily interact with each other as this enables the system to distribute its total energy among the degrees of freedom according to equipartition. The interactions between particles lead to the diffusion of correlations, which is described by $\partial_t\left\langle \mathcal{O}(t)\mathcal{O}(t') \right\rangle_{\mathrm{eq}}$ in \eref{Kubo}. If the system is slightly perturbed away from equilibrium, the energy of the system will be redistributed towards equipartition due to the interactions in the system. This will lead to a dissipation of the perturbation and a relaxation back to equilibrium, which is described by the left hand side of equation \eref{Kubo}. Thus, the response of the system $\chi(t, t')$ and the diffusion $\partial_t\left\langle \mathcal{O}(t)\mathcal{O}(t') \right\rangle_{\mathrm{eq}}$ have the same physical origin, viz.\ the momentum transfer between the microscopic degrees of freedom.

The FDRs in KFT are derived within the free theory far away from equilibrium, thus the physical picture is different. Within a single realization of the initial conditions, the particle $j$ has a constant momentum $\vec{p}_j^{\,(\mathrm{i})}$. However, in the ensemble seen as a whole, the momentum of particle $j$ is random with velocity variance $\sigma_1^2/3$ due to the averaging over all realizations.

The solutions to the free equations of motion with inhomogeneity $\mathbf{K}$, see \eref{solution_position}, describe the propagation of this random initial momentum in terms of the propagator $g_{qp}$. However, the propagation of the inhomogeneity is described by $g_{qp}$ as well. The physical origin of the FDRs in KFT could thus be seen as a consequence of Newton's 2nd axiom: The inhomogeneity changes the momentum by an amount which then has to propagate like a momentum. The FDRs in KFT are thus purely kinematic arguments which makes them, however, not less valuable. In our derivation of the time-reversal symmetry we crucially used the Gaussian form of the initial conditions, so these kinematic FDRs seem to rely on the special form of the initial conditions.

In thermal systems, FDRs describe the linear response of the system to small departures from equilibrium. In KFT we are dealing with a system far from equilibrium and consider departures from the free theory. Keep in mind that the `free' theory already contains some interactions since we use the Zel'dovich propagator, cf.\ \eref{splitting_Hamiltonian}. Through the time-reversal symmetry we were able to prove a whole hierarchy of higher-order FDRs which means that our interpretation of FDRs is not limited to linear departures from the free theory.

To our knowledge this type of kinematic FDRs far from equilibrium is a novel relation and is of significant interest because it can describe systems far away from equilibrium.

\section{Conclusion and Outlook}
\label{sec:Conclusion}

In the context of the recently developed formalism of KFT \cite{Bartelmann_main, Bartelmann_factorization} we have shown that the process of cosmic structure formation can be split into three processes: Particle diffusion due to the initial momentum variance $\sigma_1^2/3$, accumulation of structure due to the initial conditional probability $C_{p_j p_k}$ between the momenta of two particles, and interactions relative to the inertial evolution.

We observed that the processes of diffusion and accumulation of structure are delicately balanced and for late times result in a net diffusion. The delicate balance is a consequence of the fact that our formalism allows to take the full non-linear coupling of free trajectories by initial momentum correlations into account and so far explicitly relies on the Gaussian form of the initial conditions. 

Including the contributions from interactions in the Born approximation, we were able to compute the time derivative of the full non-linear density power spectrum and compare the Born approximation with simulations. We observed that the naive intuition, coming from perturbative approaches to cosmic structure formation, that the relative error of the approximation should be smaller at high redshifts and large scales, is not valid here as our approach is non-perturbative, but approximative. At low redshift, the relative error is of order $\lesssim\SI{30}{\percent}$ even on highly non-linear scales up to $k\sim 5\, h\,\mathrm{Mpc}^{-1}$. We postpone a thorough analysis of the time evolution of the relative error to a future study.

We observed a tendency of the net diffusion to approach the interaction contribution over time suggesting that diffusion and interaction are kinematically related to each other. The FDRs discussed in Section \ref{sec:FDR} show that the evolution of diffusion is indeed kinematically related to the reaction of the system to a gradient force. This result is independent of the form of the gravitational potential.

Although KFT describes a system far from equilibrium, we find that kinematic FDRs hold and in fact a whole hierarchy of higher order FDRs follows from a time-reversal symmetry of the generating functional. The Gaussian form of the initial conditions seems to be crucial for the validity of the FDRs as well as the form \eref{solution_position} of the trajectories showing that the initial momentum is propagated in the same way as an inhomogeneity.

Finally, we want to give a rather speculative outlook into future applications and relevance of the FDRs found in this work. It is yet unclear whether the FDRs are somehow related to the virial theorem, both being derived from kinematic arguments and relating the interactions and inertial movement in a system. The FDRs might, thus, be used to study virialization and the stable clustering regime. In fact our results in Fig.\ \ref{fig:equilibrium} indicate that the relative sum of net diffusion and interactions drops to zero on very small scales. If this remains true when taking post-Born approximations into account, which is necessary on these scales, this would imply that an equilibrium between clustering and diffusion is reached on very small scales. Furthermore, since the Gaussian form of the initial conditions plays an important role for the validity of the FDRs and the cancellation of the one- and two-particle contributions in Fig.\ \ref{fig:balance}, FDRs might be used in future studies to examine how non-Gaussianities effect the formation of a stable-clustering regime.

\ack

We would like to thank Shankar P. Das for recommending us to study fluctuation-dissipation relations in KFT. We are also grateful to Barbara Drossel for a discussion on fluctuation-dissipation relations in general. This work was supported in part by the Transregional Collaborative Research Centre TRR 33 `The Dark Universe' of the German Science Foundation. Futhermore, JD would like to thank the cosmology group at the ITA Heidelberg, amongst others Robert Lilow, Carsten Littek, Elena Kozlikin, Sara Konrad and Julius Mildenberger, for many important discussions and the opportunity to work in such a great environment.

\appendix

\section{Proof of the time-reversal symmetry}
\label{Time_reversal_prove}

We search here for the time-reversal transformation that leaves the free generating functional invariant assuming that the momentum part of $\mathbf{J}$ and the spatial part of $\mathbf{K}$ vanish. We make the following ansatz for the transformation:
\begin{eqnarray}
\mathcal{T}\vec{q}_j(t)&=\vec{q}_j(-t), \\
\mathcal{T}\vec{p}_j(t)&=-\vec{p}_j(-t), \\
\mathcal{T}\vec{\chi}_{q_j}(t)&=-\vec{\chi}_{q_j}(-t), \\
\mathcal{T}\vec{\chi}_{p_j}(t)&=\vec{\chi}_{p_j}(-t)-\mathrm{i}C^{-1}_{p_jp_k}\vec{p}_k(-t) -\vec{c}_j(-t),
\end{eqnarray}
where $\vec{c}_j(t)$ remains to be determined. In order to prove invariance of the generating functional we first have to check how the MSR-action, determining the dynamics of the system, changes under the transformation. After we have shown how the dynamics of the system transform, we use these dynamics to prove invariance of the generating functional and determine the appropriate form of the remaining free parameter $\vec{c}_j(t)$ of the transformation.

\subsection{Transformation of the MSR-action}

The dynamics of the free generating functional are determined by the free MSR-action:
\begin{equation}
S_0[\mathbf{x}, \boldsymbol{\chi}]= \int\mathrm{d}t\,\boldsymbol{\chi}\cdot \mathbf{E}_0(\mathbf{x}), \label{free_action}
\end{equation}
where the free Hamiltonian equations of motion are given by:
\begin{eqnarray}
\vec{E}_{0, q_j}&=\partial_t\vec{q}_j(t)-\vec{p}_j(t)=0, \\
\vec{E}_{0, p_j}&=\partial_t\vec{p}_j(t)=0.
\end{eqnarray}
Thus, the free MSR-action has the explicit form:
\begin{equation}
S_0[\mathbf{x}, \boldsymbol{\chi}]= \int\mathrm{d}t\,\left[ \vec{\chi}_{q_j}(t)\cdot\left(\partial_t\vec{q}_j(t)-\vec{p}_j(t)\right)+ \vec{\chi}_{p_j}(t)\cdot \partial_t\vec{p}_j(t)  \right]. \label{free_action_elaborate}
\end{equation}
The dynamics of the time-reversed generating functional \eref{TR_generating_func} are determined by the time-reversed MSR-action:
\begin{eqnarray}
\mathcal{T}S_0[\mathbf{x}, \boldsymbol{\chi}]=\int\mathrm{d}t\, &\left[ -\vec{\chi}_{q_j}(-t)\cdot\left( \partial_t\vec{q}_j(-t)+\vec{p}_j(-t) \right) \right. \nonumber \\
&-\left.\left(\vec{\chi}_{p_j}(-t)-\mathrm{i}C_{p_jp_k}^{-1}\vec{p}_k(-t)-\vec{c}_j(-t)\right)\cdot\partial_t\vec{p}_j(-t) \right] .
\end{eqnarray}
Substituting $t\rightarrow-t$ and using $-\partial_t=\partial_{-t}$ results in the simple form:
\begin{eqnarray}
\mathcal{T}S_0[\mathbf{x}, \boldsymbol{\chi}]=S_0[\mathbf{x}, \boldsymbol{\chi}] + \Delta S_0[\mathbf{x}],
\end{eqnarray}
where
\begin{equation}
\Delta S_0[\mathbf{x}] \coloneqq-\int\mathrm{d}t\, \left( \partial_t\vec{p}_j(t) \right)\cdot\left( \mathrm{i}C_{p_jp_k}^{-1}\vec{p}_k(t) + \vec{c}_j(t) \right).
\end{equation}
We observe that the term $\Delta S_0$ contains no $\chi$-functions and is thus unimportant for the dynamics of the system (but is nevertheless essential in the following). The rest of $\mathcal{T}S_0$ contains $\chi$-functions and therefore determines our dynamics. This part is unchanged compared to the free action \eref{free_action_elaborate}. This is a consequence of the time-reversal invariance of the Hamiltonian equations of motion. Thus, the dynamics of the system are invariant under time reversal with one caveat: Due to the change of the direction of time, causality now demands advanced instead of retarded Green's functions in the solutions \eref{solution_position} and \eref{solution_momentum} to the equations of motion:
\begin{eqnarray}
\vec{q}_j(t)&= \vec{q}_j^{\,(\mathrm{i})}+ g_{qp}(t, 0)\vec{p}_j^{\,(\mathrm{i})}-\int\mathrm{d}t'\,G^{(\mathrm{adv})}_{qp}(t, t')\vec{K}_{p_j}(t'), \label{TR:spatial_solution} \\
\vec{p}_j(t)&=\vec{p}_j^{\,(\mathrm{i})} -\int\mathrm{d}t'G^{(\mathrm{adv})}_{pp}(t, t')\vec{K}_{p_j}(t'). \label{TR:momentum_solution}
\end{eqnarray}
The advanced propagators are given by:
\begin{equation}
G^{(\mathrm{adv})}_{qp}(t, t')=-g_{qp}(t, t')\Theta(t'-t),\quad\text{and}\quad G^{(\mathrm{adv})}_{pp}(t, t')=-\Theta(t'-t). \label{Adv_Greens}
\end{equation}

\subsection{Transformation of the generating functional}

Under the assumption that the momentum part of $\mathbf{J}$ and the spatial part of $\mathbf{K}$ vanish, the free generating functional \eref{Generating_functional} has the form:
\begin{eqnarray}
\fl Z_0\big[ \mathbf{J},\mathbf{K} \big] = \int\mathrm{d}\Gamma\exp\left\{ \mathrm{i}\int\mathrm{d}t\,\vec{J}_{q_j}(t)\left[ \vec{q}_j^{\,(\mathrm{i})}+ g_{qp}(t, 0)\vec{p}_j^{\,(\mathrm{i})}-\int\mathrm{d}t'\,G^{(\mathrm{ret})}_{qp}(t, t')\vec{K}_{p_j}(t') \right] \right\}. \label{gen_functional_elaborate}
\end{eqnarray}
Having determined the dynamics for the time-reversed generating functional, we can derive the analogous form for the time-reversed generating functional using the solutions \eref{TR:spatial_solution} and \eref{TR:momentum_solution} to the equations of motion:
\begin{eqnarray}
\fl\mathcal{T}Z_0\big[ \mathbf{J},\mathbf{K} \big] &=\int\mathrm{d}\Gamma\exp\left( \mathrm{i}\int\mathrm{d}t\,\mathbf{J}\cdot \bar{\mathbf{x}} + \mathrm{i}\Delta S_0 [\bar{\mathbf{x}}] \right) \nonumber \\
\fl&=\int\mathrm{d}\Gamma\exp\left\{\mathrm{i}\int\mathrm{d}t\,\vec{J}_{q_j}(t)\left[ \vec{q}_j^{\,(\mathrm{i})}+g_{qp}(t, 0)\vec{p}_j^{\,(\mathrm{i})}-\int\mathrm{d}t'\,G_{qp}^{(\mathrm{adv})}(t, t')\vec{K}_{p_j}(t') \right] \right. \nonumber \\ 
\fl& + \left. \mathrm{i}\int\mathrm{d}t\,\vec{K}_{p_j}(t)\cdot\vec{c}_j(t)-\int\mathrm{d}t\, \vec{K}_{p_j}(t)C^{-1}_{p_jp_k}\left[ \vec{p}_k^{\,(\mathrm{i})}-\int\mathrm{d}t'\,G^{(\mathrm{adv})}_{pp}(t, t')\vec{K}_{p_k}(t') \right]\right\}. \label{App:tr_gen_func}
\end{eqnarray}
The aim of this section is to show that this expression for the time-reversed generating functional coincides with the non-transformed generating functional \eref{gen_functional_elaborate} for the correct choice of $\vec{c}_j$. We see that they already coincide for $\mathbf{K}=0$, but it remains to show that the parts containing the auxiliary source $\mathbf{K}$ also agree.

Due to their Gaussian form, the initial conditions \eref{Initial_cond} obey:
\begin{equation}
-\frac{\partial}{\partial \vec{p}_j^{\,(\mathrm{i})}} P\big( \mathbf{x}^{(\mathrm{i})} \big)= P\big( \mathbf{x}^{(\mathrm{i})} \big) C^{-1}_{p_j p_k} \vec{p}_k^{\,(\mathrm{i})}. \label{derivative_prob_distr}
\end{equation}
Using this identity we show in \ref{Exponential_expansion} that:
\begin{eqnarray}
&\exp\left( \int \mathrm{d}t' \vec{K}_{p_j}(t')\cdot\frac{\partial}{\partial \vec{p}_j^{\,(\mathrm{i})}}\right) P\big( \mathbf{x}^{(\mathrm{i})} \big) \nonumber \\
&=P\big( \mathbf{x}^{(\mathrm{i})} \big) \exp\left[ -\int\mathrm{d}t\, \vec{K}_{p_j}(t)C^{-1}_{p_jp_k}\left( \vec{p}_k^{\,(\mathrm{i})}-\int\mathrm{d}t'\,G^{(\mathrm{adv})}_{pp}(t, t')\vec{K}_{p_k}(t') \right) \right]. \label{TR:derivatives_identity}
\end{eqnarray}
The time-reversed generating functional now takes on the form:
\begin{eqnarray}
\fl\mathcal{T}Z_0\big[ \mathbf{J},\mathbf{K} \big] =&\int\mathrm{d}\mathbf{x}^{(\mathrm{i})} \exp \left\{ \mathrm{i}\int\mathrm{d}t\,\vec{J}_{q_j}(t)
\left[ \vec{q}_j^{\,(\mathrm{i})}+ g_{qp}(t, 0)\vec{p}_j^{\,(\mathrm{i})}-\int\mathrm{d}t'\,G^{(\mathrm{adv})}_{qp}(t, t')\vec{K}_{p_j}(t') \right] \right\} \nonumber \\
\fl&\times \exp \left( \mathrm{i}\int\mathrm{d}t\,\vec{K}_{p_j}(t)\cdot\vec{c}_j(t) \right) \exp\left( \int \mathrm{d}t' \vec{K}_{p_j}(t')\cdot\frac{\partial}{\partial \vec{p}_j^{\,(\mathrm{i})}}\right) P\big( \mathbf{x}^{(\mathrm{i})} \big).
\end{eqnarray}
Integration by parts in the initial momenta (boundary terms vanish since the probability distribution $P\big(\mathbf{x}^{(\mathrm{i})}\big)$ drops to zero for $\vec{p}_j\rightarrow\pm \infty$) yields\footnote{To be more precise: We have to expand the operator $\exp\left( \int\mathrm{d}t'\,\vec{K}_{p_j}(t')\partial /\partial\vec{p}_j^{\,(\mathrm{i})} \right)$ in its power series and for the term of $n$th order we have to integrate $n$ times by parts.}:
\begin{eqnarray}
\fl\mathcal{T}Z_0\big[ \mathbf{J},\mathbf{K} \big] &=\int\mathrm{d}\mathbf{x}^{(\mathrm{i})}P(\mathbf{x}^{(\mathrm{i})})\exp\left( -\int \mathrm{d}t' \vec{K}_{p_j}(t')\cdot\frac{\partial}{\partial \vec{p}_j^{\,(\mathrm{i})}}\right)\exp\left( \mathrm{i}\int\mathrm{d}t\,\vec{K}_{p_j}(t)\cdot\vec{c}_j(t) \right) \nonumber \\
\fl&\times\exp\left\{\mathrm{i}\int\mathrm{d}t\,\vec{J}_{q_j}(t)\left[ \vec{q}_j^{\,(\mathrm{i})}+ g_{qp}(t, 0)\vec{p}_j^{\,(\mathrm{i})}-\int\mathrm{d}t'\,G^{(\mathrm{adv})}_{qp}(t, t')\vec{K}_{p_j}(t') \right] \right\} \nonumber \\
\fl&=\int\mathrm{d}\Gamma\exp\left( -\mathrm{i}\int \mathrm{d}t' \vec{K}_{p_j}(t')\cdot\int\mathrm{d}t\,\vec{J}_{q_j}(t)g_{qp}(t, 0)\right) \exp \left( \mathrm{i}\int\mathrm{d}t\,\vec{K}_{p_j}(t)\cdot\vec{c}_j(t) \right)\nonumber \\
\fl&\times\exp\left\{\mathrm{i}\int\mathrm{d}t\,\vec{J}_{q_j}(t)\left[ \vec{q}_j^{\,(\mathrm{i})}+ g_{qp}(t, 0)\vec{p}_j^{\,(\mathrm{i})}-\int\mathrm{d}t'\,G^{(\mathrm{adv})}_{qp}(t, t')\vec{K}_{p_j}(t') \right] \right\},
\end{eqnarray}
where we executed the derivative with respect to the initial momenta in the last step. Careful comparison with the non-transformed generating functional \eref{gen_functional_elaborate} shows that $\vec{c}_j$ has to be a functional of the auxiliary field $\mathbf{J}$ and has the form:
\begin{eqnarray}
\vec{c}_j[\mathbf{J}, t']&= -\int\mathrm{d}t\,\vec{J}_{q_j}(t)\left( G^{(\mathrm{ret})}_{qp}(t, t')-G^{(\mathrm{adv})}_{qp}(t, t') -g_{qp}(t, 0) \right) \nonumber \\
&= -\int\mathrm{d}t\,\vec{J}_{q_j}(t) \left( g_{qp}(t, t')\Theta(t-t')+g_{qp}(t, t')\Theta(t'-t)-g_{qp}(t, 0) \right) \nonumber \\
&= -\int\mathrm{d}t\,\vec{J}_{q_j}(t)\left( g_{qp}(t, t')-g_{qp}(t, 0) \right) = g_{qp}(t', 0)\int\mathrm{d}t\,\vec{J}_{q_j}(t).
\end{eqnarray}
This finally proves the time-reversal symmetry \eref{TR_symmetry} together with $\vec{c}_j$ in the form \eref{c_term}.

\section{Useful Identities}

In this section we prove the two identities \eref{TR:derivatives_identity} and \eref{one_particle_derivative} used throughout this work.

\subsection{Derivatives of the initial phase-space distribution}
\label{Exponential_expansion}

We aim at proving the identity \eref{TR:derivatives_identity}, which we write here as:
\begin{equation}
P\big( \mathbf{x}^{(\mathrm{i})} \big) E_1 E_2 = \hat{E}_3 P\big( \mathbf{x}^{(\mathrm{i})} \big), \label{A:identity_rewritten}
\end{equation}
where we introduced the abbreviations:
\begin{eqnarray}
E_1&\coloneqq \exp\left( -\int \mathrm{d}t' \vec{K}_{p_j}(t')C^{-1}_{p_j p_k} \vec{p}_k^{\,(\mathrm{i})} \right), \label{E1} \\
E_2&\coloneqq \exp\left( \int \mathrm{d}t' \mathrm{d}t\,\vec{K}_{p_j}(t)C^{-1}_{p_j p_k} \vec{K}_{p_k}(t')G^{(\mathrm{adv})}_{pp}(t, t') \right), \label{E2} \\
\hat{E}_3&\coloneqq \exp\left( \int \mathrm{d}t' \vec{K}_{p_j}(t')\cdot\frac{\partial}{\partial \vec{p}_j^{\,(\mathrm{i})}}\right). \label{E3}
\end{eqnarray}
We denoted the third exponential with a hat in order to indicate that it is an operator acting on the initial probability distribution.

In a first step we simplify the expression for the exponential $E_2$ using that the momentum propagator $G^{(\mathrm{adv})}_{pp}$ is given by \eref{Adv_Greens}:
\begin{eqnarray}
\fl\int \mathrm{d}t' \mathrm{d}t\,\vec{K}_{p_j}(t)C^{-1}_{p_j p_k} \vec{K}_{p_k}(t')G^{(\mathrm{adv})}_{pp}(t, t')&= -\int_{-\infty}^{\infty} \mathrm{d}t' \int_{-\infty}^{t'} \mathrm{d}t\,\vec{K}_{p_j}(t)C^{-1}_{p_j p_k} \vec{K}_{p_k}(t') \nonumber \\
&= -\frac{1}{2}\int_{-\infty}^{\infty} \mathrm{d}t' \int_{-\infty}^{\infty} \mathrm{d}t \,\vec{K}_{p_j}(t)C^{-1}_{p_j p_k} \vec{K}_{p_k}(t'),
\end{eqnarray}
where we used the symmetry of $C^{-1}_{p_jp_k}$ in $j\leftrightarrow k$. Introducing another abbreviation:
\begin{equation}
\vec{\gamma}_j\coloneqq\int\mathrm{d}t \,\vec{K}_{p_j}(t),
\end{equation}
the expansions of the three exponentials \eref{E1} to \eref{E3} become:
\begin{eqnarray}
E_1&=\sum_{n=0}^{\infty} \frac{1}{n!}\left( -\vec{\gamma}_jC^{-1}_{p_j p_k}\vec{p}_k^{\,(\mathrm{i})} \right)^n= \sum_{n=0}^{\infty}E_1^{(n)},  \\
E_2&=\sum_{n=0}^{\infty} \frac{2^{-n}}{n!}\left( -\vec{\gamma}_jC^{-1}_{p_j p_k}\vec{\gamma}_k \right)^{n} =\sum_{n=0}^{\infty}E_1^{(2n)}, \\
\hat{E}_3&=\sum_{n=0}^{\infty} \frac{1}{n!}\left( \vec{\gamma}_j\cdot\frac{\partial}{\partial\vec{p}_j^{\,(\mathrm{i})}} \right)^n= \sum_{n=0}^{\infty}\hat{E}_3^{(n)},
\end{eqnarray}
where an upper index $(n)$ denotes the order of a term in the auxiliary field $\mathbf{K}$ or, equivalently, in $\vec{\gamma}$. In this notation, the property \eref{derivative_prob_distr} of the initial probability distribution has the form:
\begin{equation}
\hat{E}_3^{(1)}P\big( \mathbf{x}^{(\mathrm{i})} \big)=-P\big( \mathbf{x}^{(\mathrm{i})} \big)\vec{\gamma}_jC_{p_jp_k}^{-1}\vec{p}_k^{\,(\mathrm{i})}. \label{A:derivative_identity}
\end{equation}
In \eref{A:identity_rewritten} the two exponentials $E_1$ and $E_2$ are multiplied with each other. Thus, we introduce $E\coloneqq E_1E_2$ and expand this exponential in orders of $\vec{\gamma}$ as well. We have to distinguish here between even and odd orders in $\vec{\gamma}$:
\begin{eqnarray}
E^{(2n)}&=\sum_{m=0}^{n} E_1^{(2m)}E_2^{(2(n-m))} \nonumber \\ &=\sum_{m=0}^{n}\frac{2^{m-n}}{(2m)!(n-m)!} 
\left( \vec{\gamma}_jC^{-1}_{p_j p_k}\vec{p}_k^{\,(\mathrm{i})} \right)^{2m} \left( -\vec{\gamma}_rC^{-1}_{p_r p_s}\vec{\gamma}_s \right)^{n-m}, \label{A:even_E} \\ 
E^{(2n+1)}&=\sum_{m=0}^{n} E_1^{(2m+1)}E_2^{(2(n-m))} \nonumber \\
&= -\sum_{m=0}^{n}\frac{2^{m-n}}{(2m+1)!(n-m)!} 
\left( \vec{\gamma}_jC^{-1}_{p_j p_k}\vec{p}_k^{\,(\mathrm{i})} \right)^{2m+1} \left( -\vec{\gamma}_rC^{-1}_{p_r p_s}\vec{\gamma}_s \right)^{n-m}. \label{A:odd_E}
\end{eqnarray}
Using \eref{A:derivative_identity}, it is straightforward to show by induction for any order $n$:
\begin{equation}
E^{(n)}P\big( \mathbf{x}^{(\mathrm{i})} \big)= \hat{E}_3^{(n)}P\big( \mathbf{x}^{(\mathrm{i})} \big). \label{A:decomposed_identity}
\end{equation}
This finally proves the relation \eref{A:identity_rewritten} to all orders in $\vec{\gamma}$.

\subsection{The one-particle time derivative}
\label{sec:one_particle_time_derivative}

Now we show the identity \eref{one_particle_derivative}. The correlation on the left-hand side of \eref{one_particle_derivative} contains an arbitrary number of additional density fields and thus has the form:
\begin{eqnarray}
\left\langle \vec{k}_1C_{p_jp_k}^{-1}\vec{p}_k(t_1)\rho_j(1)\dots \right\rangle &= \int\mathrm{d}\Gamma\, \vec{k}_1C_{p_jp_k}^{-1}\vec{p}_k^{\,(\mathrm{i})}\mathrm{e}^{\mathrm{i}\left(\vec{L}_{q_r}\vec{q}_r^{\;\!(\mathrm{i})}+\vec{L}_{p_r}\vec{p}_r^{\;\!(\mathrm{i})}\right)} \nonumber \\
&= \int\mathrm{d}\Gamma\,\vec{k}_1C_{p_jp_k}^{-1} \frac{\partial}{\mathrm{i}\partial\vec{L}_{p_k}}\mathrm{e}^{\mathrm{i}\left(\vec{L}_{q_r}\vec{q}_r^{\;\!(\mathrm{i})}+\vec{L}_{p_r}\vec{p}_r^{\;\!(\mathrm{i})}\right)},
\end{eqnarray}
where we used that $\vec{p}_k(t_1)=\vec{p}_k^{\,(\mathrm{i})}$ since no response fields are applied. In a final step we evaluate the integrals in the initial momenta:
\begin{eqnarray}
\left\langle \vec{k}_1C_{p_jp_k}^{-1}\vec{p}_k(t_1)\rho_j(1)\dots \right\rangle &=V^{-N}\int\mathrm{d}\mathbf{q}^{(\mathrm{i})}\,\vec{k}_1C_{p_jp_k}^{-1} \frac{\partial}{\mathrm{i}\partial\vec{L}_{p_k}} \mathrm{e}^{-\frac{1}{2}\vec{L}_{p_l}C_{p_l p_m}\vec{L}_{p_m}} \mathrm{e}^{\mathrm{i}\vec{L}_{q_r}\vec{q}_r^{\;\!(\mathrm{i})}} \nonumber \\
&=\mathrm{i}V^{-N}\int\mathrm{d}\mathbf{q}^{(\mathrm{i})}\,\vec{k}_1C_{p_jp_k}^{-1} C_{p_kp_s}\vec{L}_{p_s} \mathrm{e}^{-\frac{1}{2}\vec{L}_{p_l}C_{p_l p_m}\vec{L}_{p_m}} \mathrm{e}^{\mathrm{i}\vec{L}_{q_r}\vec{q}_r^{\;\!(\mathrm{i})}} \nonumber \\
&=\mathrm{i}\vec{k}_1\vec{L}_{p_j}Z_0\big[ \mathbf{L}, 0 \big]=\mathrm{i}\frac{3}{\sigma_1^2}D_{t_1}^{(1, j)}Z_0\big[ \mathbf{L}, 0 \big],
\end{eqnarray}
where we used $C_{p_jp_k}^{-1} C_{p_kp_s}= \delta_{js}\mathbbm{1}_3$.

\section*{References}

\bibliography{paper}

\end{document}